\documentclass[12pt]{article}
\pdfoutput=1
\usepackage{comment}
\usepackage{amsmath}
\usepackage{amssymb}
\usepackage{nccmath}
\usepackage{graphicx}
\usepackage{animate}
\usepackage{here}
\usepackage{subcaption}
\usepackage{url}
\usepackage[sort&compress, numbers, merge]{natbib}
\usepackage{braket}
\usepackage[italicdiff]{physics}
\usepackage{tikz}
\usepackage[compat=1.1.0]{tikz-feynhand}
\usepackage{slashed}
\usepackage{mathtools}
\usepackage{bbold}
\usepackage{pgfplots}
\usepackage{tcolorbox}

\tcbuselibrary{raster,skins}
\tcbuselibrary{theorems}

\usepackage{pgfplots}  
\usetikzlibrary{arrows.meta,arrows,angles,calc,perspective} 

\usepackage{tikz-3dplot}
 \tdplotsetmaincoords{60}{115}
\pgfplotsset{compat=newest}

\usetikzlibrary{3d}
\usetikzlibrary{calc}
\pgfplotsset{compat=newest}

\usetikzlibrary{decorations.markings}

\usepackage{todonotes}
\newcommand{\uone}{\mathrm{U}(1)}
\newcommand{\SU}{\mathrm{SU}}
\newcommand{\vone}{v_{\uone}}

\newcommand{\numrange}[2]{{#1}\text{--}{#2}}
\newcommand{\hc}{\text{h.c.}}

\newcommand{\gE}{\gamma_{\mathrm{E}}}

\newcommand{\temit}{t}
\newcommand{\femit}{f_\mathrm{emit}}

\newcommand{\version}{arXiv2}
\DeclareRobustCommand{\modifiedat}[2]{%
\ifthenelse{\equal{\version}{#1}}{\textcolor{red}{#2}}{#2}%
}
\newcommand{\kdot}{\mathbin{\ast}}
\numberwithin{equation}{section}

\usepackage[colorlinks=true, linkcolor=blue, citecolor=blue,
urlcolor=black]{hyperref} 
\usepackage[capitalize]{cleveref}
\crefformat{section}{Sec.\,#2#1#3}
\Crefformat{section}{Section~#2#1#3}
\crefformat{equation}{Eq.\,(#2#1#3)}
\Crefformat{equation}{Equation~(#2#1#3)}
\crefformat{figure}{Fig.\,#2#1#3}
\Crefformat{figure}{Figure~#2#1#3}

 \DeclareSymbolFontAlphabet{\mathbb}{AMSb}                
\begin{document}
\def\ps{\mathbf{p}}
\def\PS{\mathbf{P}}
\baselineskip 0.6cm
\def\simgt{\mathrel{\lower2.5pt\vbox{\lineskip=0pt\baselineskip=0pt
           \hbox{$>$}\hbox{$\sim$}}}}
\def\simlt{\mathrel{\lower2.5pt\vbox{\lineskip=0pt\baselineskip=0pt
           \hbox{$<$}\hbox{$\sim$}}}}
\def\simprop{\mathrel{\lower3.0pt\vbox{\lineskip=1.0pt\baselineskip=0pt
             \hbox{$\propto$}\hbox{$\sim$}}}}
\def\tr{\mathop{\rm tr}}
\def\SU{\mathop{\rm SU}}

\def\uoneG{\mathrm{U}(1)_{G}}
\def\VG{V}
\def\VH{v}
\def\azimuth{\theta}
\newcommand{\az}{\azimuth}
\def\zenith{\theta_s}
\def\higgsprofile{\xi}
\def\Newton{G_{\mathrm{N}}}

\begin{titlepage}

\begin{flushright}
IPMU24-0043
\end{flushright}

\vskip 1.1cm

{\centering
{\Large \bf 
Gravitational Waves from Metastable Cosmic Strings\\ in Supersymmetric New Inflation Model}

\vskip 1.2cm
Akifumi Chitose$^{a}$, 
Masahiro Ibe$^{a,b}$,
Shunsuke Neda$^{a}$ and
Satoshi Shirai$^{b}$
\vskip 0.5cm

{\it

$^a$ {ICRR, The University of Tokyo, Kashiwa, Chiba 277-8582, Japan}

$^b$ {Kavli Institute for the Physics and Mathematics of the Universe
 (WPI), \\The University of Tokyo Institutes for Advanced Study, \\ The
 University of Tokyo, Kashiwa 277-8583, Japan}

}
}

\begin{abstract}
Recent observations by pulsar timing arrays (PTAs) indicate a potential detection of a stochastic gravitational wave (GW) background. Metastable cosmic strings have been recognized as a possible source of the observed signals.
In this paper, we propose an $R$-invariant supersymmetric new inflation model.
It is characterized by a two-step symmetry breaking $\mathrm{SU}(2) \to \mathrm{U}(1)_G \to \mathrm{nothing}$, incorporating metastable cosmic strings. The field responsible for the initial symmetry breaking acts as the inflaton, while the second symmetry breaking occurs post-inflation, ensuring the formation of the cosmic string network without monopole production.
Our model predicts symmetry breaking scales consistent with the string tensions favored by PTA data, $G_\mathrm{N} \mu_\mathrm{str} \sim 10^{-5}$, where $G_\mathrm{N}$ is the Newton constant. 
Notably, a low reheating temperature is required to suppress non-thermal gravitino production from the decay of inflaton sector fields. This also helps evading LIGO-Virgo-KAGRA constraints, while yielding a distinctive GW signature that future PTA and interferometer experiments can detect.
Additionally, we examine the consistency of this scenario with non-thermal leptogenesis and supersymmetric dark matter. 
\end{abstract}
\end{titlepage}
\newpage

\tableofcontents

\section{Introduction}
Recently, pulsar timing array (PTA) collaborations reported an observation of the Hellings-Downs angular correlation in the nanohertz frequency range, suggesting evidence for a stochastic gravitational wave (GW) signal~\cite{NANOGrav:2023gor,EPTA:2023fyk,Reardon:2023gzh,Xu:2023wog}. 
From the perspective of new physics searches, stochastic GWs have served as a vital observational channel for investigating cosmic strings (for reviews, see, e.g., Refs.\,\cite{Vilenkin:1982hm,Hindmarsh:2011qj,Auclair:2019wcv,Gouttenoire:2019kij}). 
Cosmic strings are topological defects that emerge from a $\uone$ symmetry breaking, with their stability ensured by the non-trivial topological charge  $\pi_1\qty[\mathrm{U}(1)] = \mathbb{Z}$.
These strings form a network that exhibits scaling behavior, maintaining a correlation length on the order of the Hubble length at all times while generating loops through reconnection. The resulting cosmic string loops gradually contract by emitting GWs until they eventually vanish, thereby contributing to the stochastic GW background.

Interestingly, the observed GW spectrum appears to favor a metastable cosmic string network rather than a stable one~\cite{NANOGrav:2023hvm}. 
In the case of metastable cosmic strings, they decay in the later Universe, causing a suppression in the low-frequency GW spectrum (see Refs.~\cite{Leblond:2009fq,Buchmuller:2019gfy,Buchmuller:2020lbh,Buchmuller:2021mbb,Buchmuller:2023aus} for further details). This suppression in the low-frequency region provides a better fit to the PTA signal than a stable string network. 
These recent PTA observations have thus sparked considerable interest in metastable cosmic strings.

Metastable strings are predicted in models that undergo a two-step symmetry breaking, such as:
\begin{align}
G \rightarrow \mathrm{U}(1) \rightarrow \mathrm{nothing} \ ,
\end{align}
where the vacuum expectation values (VEVs) are hierarchical. 
In such models, $\pi_1\qty[G]$ is trivial while $\pi_2\qty[G/\mathrm{U}(1)] = \pi_1[\mathrm{U}(1)] = \mathbb{Z}$. 
In this setup, cosmic strings emerge as classically stable configurations due to the non-trivial topological charge of $\pi_1[\mathrm{U}(1)]$ in the low-energy effective theory below the $G \to \mathrm{U}(1)$ symmetry-breaking scale. 
However, in the full theory, cosmic strings are ultimately metastable since $\pi_1[G] = 0$, meaning no topological charge protects them. 
More closely, these models feature monopole configurations associated with the topological charge $\pi_2[G/\mathrm{U}(1)]$ at the first symmetry breaking. The breaking of the strings can be understood as the Schwinger effect~\cite{Preskill:1992ck}, which generates monopole-antimonopole pairs within the string, leading to the fragmentation of long strings into smaller pieces.

Notably, the string breaking rate inferred from the PTA signal suggests that the mass of the magnetic monopole $m_M$ (of order of the first symmetry breaking scale) 
and the cosmic string tension $\mu_\mathrm{str}$ (of order of the second symmetry breaking scale) 
are not hierarchically separated, i.e., $m_M \sim \sqrt{\mu_\mathrm{str}}$. 
This raises questions about whether the symmetry breaking occurs in a sequential manner that allows for the formation of a long cosmic string network as anticipated above. 
It is also unclear whether the monopoles generated at the first symmetry-breaking stage are sufficiently diluted to avoid the monopole problem.

In light of the PTA observations, there has been significant discussion focusing on realistic models and cosmological scenarios that lead to the formation of metastable strings, particularly within the framework of supersymmetric Grand Unified Theories (GUTs)~\cite{Buchmuller:2021dtt,Buchmuller:2024zzk,Antusch:2023zjk,Antusch:2024nqg,Ahmed:2023pjl,Ahmed:2024iyd,Lazarides:2023rqf,Afzal:2023cyp,Pallis:2024mip,Maji:2024cwv}. 
There are clear reasons for this interest. First, the PTA signal not only suggests that the scales of the two-step symmetry breaking are close to each other but also favors scales around the GUT scale, i.e., $10^{\numrange{15}{16}}$ GeV. Furthermore, the mechanism of GUT symmetry breaking is highly compatible with hybrid inflation, leading to the proposal of various models incorporating this framework.
In such scenarios, however, both the monopole-generating symmetry breaking and the U(1) symmetry breaking occur after inflation, necessitating further investigation of the monopoles and the cosmic string network formation processes.

In this paper, we propose an alternative inflationary scenario that accommodates a successful picture of metastable cosmic strings, where the first symmetry-breaking field (i.e., the $G$-breaking field) serves as the inflaton (see Refs.\,\cite{Pallis:2024joc,Antusch:2024qpb}). 
Specifically, we focus on an $R$-symmetric supersymmetric new inflation model~\cite{Izawa:1996dv,Izawa:1997df}. 
An advantage of this type of model is that the symmetry enhancement point is naturally selected as the initial condition through pre-inflationary dynamics~\cite{Izawa:1997df}.
We consider the two-step symmetry breaking of $G=\SU(2)$ gauge symmetry to $\uone$ by an $\SU(2)$ adjoint field, which plays the role of the inflaton in new inflation model. 
The remaining $\uone$ symmetry is then broken by $\SU(2)$ doublet fields~\cite{Kephart:1995cg}. 
In this setup, $\SU(2)$ symmetry is broken at the onset of inflation, whereas $\uone$ symmetry breaking occurs post-inflation. As a result, the cosmic string network is formed after inflation while monopoles are diluted during inflation, ensuring they do not interfere with the formation of the cosmic string network.

As a characteristic feature,
the model predicts 
a string tension of order of the GUT scale. 
Besides, it turns out that the model requires a low reheating temperature after inflation to avoid excessive gravitino production from the decay of inflaton sector fields. 
Interestingly, this low reheating temperature enables the GW spectrum to be consistent with the constraints set by the LIGO–Virgo–KAGRA (LVK) collaboration~\cite{KAGRA:2021kbb}, even for a relatively high string tension at the GUT scale, i.e., $G_{\mathrm{N}} \mu_\mathrm{str} \sim 10^{-5}$, where $G_{\mathrm{N}}$ denotes the Newton constant.

We also provide a working example for a successful cosmology,
where the baryon asymmetry of the Universe 
is provided by the non-thermal leptogenesis and the dark matter is provided by the lightest supersymmetric particle (LSP). 
We demonstrate that these scenarios, along with the present inflation model, are all consistent within the framework of a high-scale supersymmetry model featuring PeV-scale gravitino/sfermion masses and TeV-scale anomaly-mediated gaugino masses.

The organization of this paper is as follows: In \cref{sec:Model}, we summarize the model of supersymmetric $R$-invariant new inflation, where the symmetry-breaking field acts as the inflaton. 
In \cref{sec:inflation}, we analyze the model, determining the model parameters that reproduce the observed curvature perturbations and spectral index of the cosmic microwave background (CMB) anisotropies. 
In \cref{sec:reheating}, we discuss the 
post inflation dynamics and reheating process addressing the gravitino problem, dark matter, and non-thermal leptogenesis. In \cref{sec:CSGWS}, we explore the properties of cosmic strings and present the predicted GW spectrum. Finally, \cref{sec:conclusions} 
concludes with discussions.

\section{Model}
\label{sec:Model}
\subsection{Symmetry Breaking  Pattern and Summary of Model}
Before discussing the details,
let us summarize features of the model.
We consider a supersymmetric 
$\SU(2)$ gauge theory
where $\SU(2)$ is 
broken down to a $\uone$ gauge symmetry
by an $\SU(2)$
adjoint 
chiral superfield $\phi^a$
$(a=1,2,3)$.
We refer to this residual $\uone$ gauge symmetry as the $\uoneG$ symmetry.
The residual $\uoneG$ gauge symmetry is broken by a pair of doublet chiral superfields $H=(H_1,H_2)$ and $\bar{H}=(\bar{H}_1,\bar{H}_2)$.%
\footnote{The indices of the triplet are occasionally interchanged, i.e., $\phi^a = \phi_a$.}
Hereafter, we assume the VEVs of these fields are given by 
\begin{align}
\expval{\phi_3}=v_{\SU(2)}  \ , \quad 
\expval{H_1} = \expval{\bar{H}_1} = v_{\mathrm{U}(1)}\ .
\end{align}
(see \cref{eq:VEVs}),
where both $v_\mathrm{U(1)}$
and $v_\mathrm{SU(2)}$ are of $\order{10^{16}}$\,GeV.
In the following, we construct a model of new inflation where 
$\phi_3$ plays the role of the inflaton.

We assume a gravitino mass in the PeV range, where supersymmetry breaking occurs in a dynamical sector separated from the inflaton sector.
As we will see, the Hubble parameter during inflation is $H_\mathrm{inf}=\order{10^{10}}$\,GeV. 
The dynamical scale of the supersymmetry breaking is set to $\order{10^{\numrange{12}{13}}}$\,GeV for PeV-scale gravitino mass, and thus, we assume it has occurred before the onset of the inflation. 
We will also see that 
the soft supersymmetry-breaking masses are smaller than the supersymmetric masses in the inflaton sector, and hence, 
the soft masses do not affect the inflation dynamics.
Note also that the energy density of the supersymmetry-breaking sector to the inflation is subdominant.

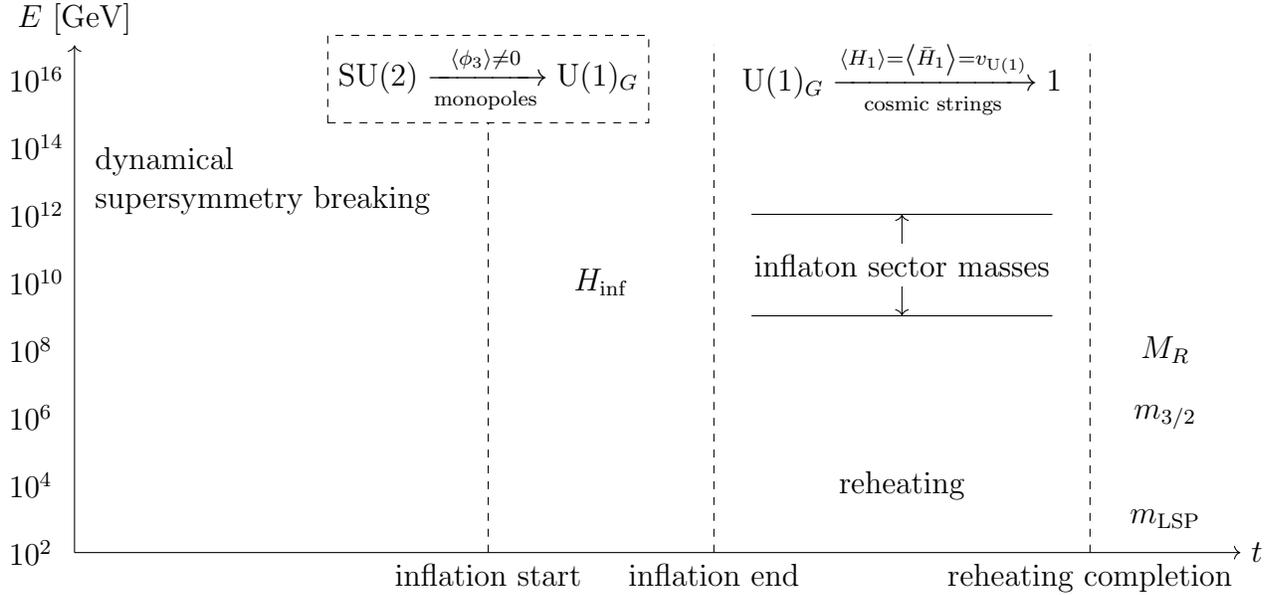
\begin{figure}[t]
    \centering
        
    \newcommand{\tSUSY}{1.5}
    \newcommand{\treh}{11}
    \newcommand{\tLSP}{14.5}
    \newcommand{\tinfl}{5.5}
    \ExplSyntaxOn
    \newcommand{\powerat}[1]{
        \int_set:Nn \l_tmpa_int { 2 * (1 + #1) }
        \int_use:N \l_tmpa_int
        \int_compare:nNnTF { \l_tmpa_int } < { 10 }
            { \phantom{9} } { }
    }
    \ExplSyntaxOff
    \newcommand{\partition}[2]{
        \draw[dashed] (#1, 0) node [below, align=center] {#2} -- (#1, 7.5)
    }
    
    \begin{tikzpicture}[y=0.9cm]
        \draw[->] (0, 0) -- (0, 7.5) node[above] {$E$ [GeV]};
        \draw[->] (0, 0) -- (15.5, 0) node[right] {$t$};
        \foreach \y in {0, ..., 7} {
          \node[left] at (0, \y) {$10^{\powerat{\y}}$};
        }
        \node[align=left] at (2.5, 5.5) {
        dynamical \\ supersymmetry  breaking};
        \partition{\tinfl}{inflation start};
        \partition{8.5}{inflation end};
        \partition{13.5}{reheating completion};
        \node[fill=white, dashed,draw] at (\tinfl, 7) {$\SU(2)\xrightarrow[\text{monopoles}]{\expval{\phi_3}\ne 0}\uoneG$};
        \node at (7, 4) {$H_{\mathrm{inf}}$};
        
        \node at (\treh, 7) {$\uoneG\xrightarrow[\text{cosmic strings}]{\expval{H_1}=\expval{\bar{H}_1}=v_{\mathrm{U}(1)}}1$};
        \draw[<->] (\treh, 3.5) -- (\treh, 5);
        \draw ($(\treh, 3.5)-(2, 0)$) -- ($(\treh, 3.5)+(2, 0)$);
        \draw ($(\treh, 5)-(2, 0)$) -- ($(\treh, 5)+(2, 0)$);
        \node[fill=white] at (\treh, 4.25) {inflaton sector masses};
        \node at (\treh, 1) {reheating};
        
        \node at (\tLSP, 0.5) {$m_{\mathrm{LSP}}$};
        \node at (\tLSP, 2) {$m_{3/2}$};
        \node at (\tLSP, 3) {$M_R$};
    \end{tikzpicture}
    
    \caption{The timeline of the events and relevant energy scales in the present scenario.
    The mass parameters $M_R$, 
    $m_{3/2}$, $m_\mathrm{LSP}$ 
    are the right-handed neutrino mass, the gravitino mass, and the LSP mass, respectively.
    Since the $\SU(2)$ breaking field drives the inflation, the monopoles are diluted away.
    \label{fig:timeline}}
\end{figure}

\Cref{fig:timeline} shows the relevant energy scales and the timeline of the events.
As we mentioned above, the supersymmetry breaking has taken place before inflation.
At the end of inflation the adjoint field obtains 
the VEV, $v_{\mathrm{SU}(2)}$,
and the doublets obtain the VEV, 
$v_{\mathrm{U}(1)}$, which are both of $\order{10^{16}}$\,GeV.
The masses of the inflaton sector fields are of $\order{10^{\numrange{10}{12}}}$\,GeV (see \cref{sec:reference}).
The reheating temperature is 
low: $T_R = \order{10^{\numrange{3}{4}}}$\,GeV
(see \cref{sec:gravitino}).

\subsection{PTA GW Signal and Breaking Scales}
\label{sec:PTA}
To estimate the symmetry breaking scale of interest, we summarize the interpretation of the PTA signal from gravitational waves produced by a metastable cosmic string network. Based on the strength of the observed spectrum, the favored string tension $\mu_\mathrm{str}$ is within the range
\begin{align}
\label{eq:Gmu range}
-6 \lesssim \log_{10} G_{\mathrm{N}} \mu_\mathrm{str} \lesssim -4
\end{align}
(see Ref.\,\cite{NANOGrav:2023hvm}). 
It is important to note that this range is derived by fixing certain parameters, such as those related to the velocity-dependent one-scale (VOS) model for the string loop density and the coefficient of the GW power spectrum. Due to theoretical uncertainties, the precise value of $G_{\mathrm{N}} \mu_\mathrm{str}$ remains highly uncertain. Thus, this parameter range should be treated as a reference rather than a definitive constraint.

In terms of the $\uoneG$ breaking scale, the string tension is given by
\begin{align}
\label{eq:tension}
    \mu_\mathrm{str} = 2\pi f_T \times v_\mathrm{U(1)}^2 \
     ,
\end{align}
where $f_T$ is a dimensionless factor which depends on model parameters.
In \cref{sec:CosmicString}, we find that typically
\begin{align}
    f_T \simeq 0.1\mbox{--} 0.2  \ , 
\end{align}
for the benchmark scenarios discussed later.
The constraint \eqref{eq:Gmu range} is translated to
\begin{align}
\label{eq:U1range}
5 \times 10^{15}\,\mathrm{GeV}  \lesssim f_T^{1/2} v_\mathrm{U(1)}\lesssim 5\times 10^{16}\,\mathrm{GeV}\ .
\end{align}

Metastable cosmic strings decay via monopole-antimonopole pair production.
The breaking rate per unit length is conventionally given by the breaking parameter $\kappa$~\cite{NANOGrav:2023hvm}, \begin{align}
    \Gamma_d= \frac{\mu_\mathrm{str}}{2\pi}e^{-\pi \kappa}\ .
\end{align}
The observed GW spectrum favors
\begin{align}
\sqrt{\kappa}\sim 8\ .
\end{align}

Theoretically, $\sqrt{\kappa}$ is approximately given by~\cite{Preskill:1992ck}
\begin{align}
    \label{eq:rootkappaPreskill}
    \sqrt{\kappa} \simeq \frac{m_M}{\sqrt{\mu_\mathrm{str}}}\ .
\end{align} 
Note that 
this approximation
is valid only for $m_M/\sqrt{\mu_\mathrm{str}} \gg 10$--$20$~\cite{Chitose:2023dam}. 
We do not explore this issue further and introduce an unknown correction factor $r_\kappa$, 
\begin{align}
\label{eq:rootkappa}
    \sqrt{\kappa} = \frac{m_M}{r_\kappa\sqrt{\mu_\mathrm{str}}}\ ,
\end{align}
and take $r_\kappa=1$--$3$ as a reference value.

The monopole mass is given in terms of the SU(2) breaking scale as
\begin{align}
\label{eq:monopolemass}
    m_M = \frac{4\pi f_M v_\mathrm{SU(2)}}{g}\ ,
\end{align}
where $g$ is the $\SU(2)$ gauge coupling and $f_M\ge 1$ is an $\order{1}$ coefficient.%
\footnote{In the 
Bogomol'nyi-Prasad-Sommerfield (BPS) 
limit~
\cite{Bogomolny:1975de,Prasad:1975kr}, the coefficient becomes $f_M=1$.}
By substituting \cref{eq:monopolemass} into \cref{eq:rootkappa}, we find that the appropriate range of the SU(2) breaking scale is 
\begin{align}
  \frac{f_M v_\mathrm{SU(2)}}{g} \simeq \sqrt{\frac{\kappa}{8\pi}}\times r_\kappa f_T^{1/2} v_\mathrm{U(1)}\ ,
\end{align}
and hence,
\begin{align}
\label{eq:SU2range}
5.7 \times 10^{15}\,\mathrm{GeV}   \lesssim \frac{f_M v_\mathrm{SU(2)}}{gr_\kappa} \lesssim 5.6\times 10^{16}\,\mathrm{GeV}\ ,
\end{align}
for $\sqrt{\kappa}\simeq 8$.
On those grounds, we take both $v_{\uone}$ and $v_{\SU(2)}$ to be $\order{10^{16}}$\,GeV in the following.

From
\cref{eq:U1range} and 
\cref{eq:SU2range},
we find $g=\order{1}$
to achieve $v_{\mathrm{\SU(2)}}\sim v_{\mathrm{\uone}}$.
Since this is not very small, we need to ensure that the SU(2) dynamical scale is smaller than the Hubble scale during inflation, so that the subsequent perturbative analysis remains valid.
The SU(2) dynamical scale is given by
\begin{align}    \Lambda_\mathrm{SU(2)} = v_\mathrm{SU(2)} \times \exp\qty[-\frac{8\pi^2}{b_{\mathrm{SU(2)}}}\frac{1}{g^2(v_\mathrm{SU(2)})}]\ , 
\end{align}
where $b_\mathrm{SU(2)}=3$ 
is the coefficient of the one-loop beta function, 
and we take $v_\mathrm{SU(2)}$ as the initial condition of the renormalization group running.
For $g(v_\mathrm{SU(2)})=1$, for example, the exponential factor 
is about $4\times 10^{-12}$,
and hence, 
$\Lambda_{\SU(2)}\ll H_\mathrm{inf}$.
Thus, the strong dynamics 
does not affect the inflation dynamics for $g(v_\mathrm{SU(2)})\lesssim 1$.
In the following analysis, we fix $g(v_\mathrm{SU(2)})=1$.

\subsection{Superpotential and Supersymmetric Vacuum}
For the inflaton sector,
we introduce two SU(2) singlet chiral fields, $X$ and  $Y$,
and consider the following superpotential,
\begin{align}
\label{eq:Wpot}
    W = X\qty(v_X^2-\lambda_X \frac{(\phi\cdot\phi)^n}{M_\mathrm{Pl}^{2n-2}}) + Y\qty(v_Y^2-\lambda_Y\frac{(\bar{H}H)^n}{M_\mathrm{Pl}^{2n-2}}) + \zeta\qty(\bar{H}\phi^a t^aH)\frac{(\bar{H}H)^{n'}}{M_\mathrm{Pl}^{2n'}} 
   \ ,
\end{align}
where $t^a=\sigma^a/2$ are the halves of the Pauli matrices  
and $\phi\cdot \phi = \phi^a\phi^a$ with implicit summation over $a$.
The parameters  
$\lambda_X$, $\lambda_Y$
and $\zeta$ are dimensionless coupling constants, $v_X$, $v_Y$ are the parameters with mass dimension one, and $M_\mathrm{Pl}\simeq 2.435 \times 10^{18}$\,GeV is the 
reduced Planck scale.
Without loss of generality, we can take $\lambda_{X,Y}$ and $v_{X,Y}$ to be positive.
The integer $n$ will be fixed in \cref{sec:inflation}, 
while we take
$n' = n - 2$ to avoid the domain wall problem after inflation (see \cref{sec:domain wall}).

The terms involving $X$ and $Y$ induce the scalar potential 
which results in the breaking of the SU(2) symmetry. The $\zeta$-term aligns the direction of the breaking between $\phi^a$,  $H$ and $\bar{H}$.
The structure of the superpotential can be achieved, for example, by the symmetries shown in Table~\ref{tab:symmetry}.%
\footnote{Note that 
the symmetry allows terms $X^{n_X}Y^{n_Y}$ with $n_X + n_Y = 2n+1$.
We assume the coefficients of those terms are slightly suppressed, which may be explained by some symmetry and their breaking.
The presence of those terms do not affect inflation dynamics significantly as long as $X$ and $Y$ have positive Hubble mass terms during inflation. 
}
Here, the $\mathbb{Z}_{4nR}$ symmetry refers to a discrete $R$-symmetry in supersymmetric theory, where the superpotential carries a charge of $2$, and the gauginos carry a charge of $1$.
Due to the discrete $R$-symmetry, the superpotential does not have a constant term.

Note that the charges of $X$ and $Y$ are identical. Therefore, both $X$ and $Y$ may couple to $(\phi \cdot \phi)^n$ and $(\bar{H}H)^n$. Without loss of generality, we can define $X$ such that it does not couple to $(\bar{H}H)^n$, while $Y$ couples to both $(\phi \cdot \phi)^n$ and $(\bar{H}H)^n$. In this paper, we omit the coupling of $Y$ to $(\phi \cdot \phi)^n$ for simplicity, as this interaction does not result in any essential difference.
Note also that the cutoff of the higher-dimensional operators are not necessarily identical to the Planck scale, but can be a lower physical scale.
In that case, $\lambda$'s 
and $\zeta$ are not 
restricted to be of $\order{1}$ (see \cref{sec:largelam}).

\begin{table}[t]
\centering
    \begin{subtable}[t]{0.45\textwidth}
\centering
\begin{tabular}{c||c|c|c|c}
& SU(2) & $\mathbb{Z}_{4nR}$ & $\mathbb{Z}_{2n}$ & $\uone_H$ \\
\hline
$\phi^a$ & $\mathbf{3}$ & $2$ & $2$ & $0$ \\
\hline
$H$ & $\mathbf{2}$ & $-1$ & $-1$ & $-1$\\
\hline
$\bar{H}$ & $\bar{\mathbf{2}}$ & $1$ & $3$&+1 \\
\hline
$X$ & -- & $2$ &$0$ & $0$ \\
\hline
$Y$ & -- & $2$ & $0$& $0$ \\
\end{tabular}
 \end{subtable}
     \caption{Charge assignment of the chiral superfields under the SU(2) gauge symmetry and discrete $\mathbb{Z}_{4nR}$ and $\mathbb{Z}_{2n}$ symmetry.
    The $\uone_H$ symmetry is anomaly free, and hence, can be an exact symmetry (if you want).
    }
    \label{tab:symmetry}
\end{table}

Let us discuss the supersymmetric vacuum of the model.
The $F$-term vacuum conditions are given by, 
\begin{align}
   - F_X^\dagger &= \qty(v_X^{2}-\lambda_X\frac{(\phi\cdot\phi)^{n}}{M_\mathrm{Pl}^{2n-2}})=0\ ,
    \\
     -F_Y^\dagger &=     \qty(v_Y^2-\lambda_Y\frac{(\bar{H}H)^{n}}{M_\mathrm{Pl}^{2n-2}})=0 \ ,\\
     \label{eq:Fphia}
     -F_{\phi^a}^\dagger & = 
     -\frac{2n\lambda_X}{M_\mathrm{Pl}^{2n-2}}X\phi^a(\phi\cdot\phi)^{n-1} + \zeta \bar{H}t^aH
     \frac{(\bar{H}H)^{n'}}{M_{\mathrm{Pl}}^{2n'}}
     = 0  \ , \\
     -F_{H}^\dagger &=-n \lambda_Y  \frac{(\bar{H}H)^{n-1}}{M_\mathrm{Pl}^{2n-2}}Y\bar{H} + \zeta \bar{H}\phi^at^a\frac{(\bar{H}H)^{n'}}{M_{\mathrm{Pl}}^{2n'}} +
      n'\zeta \bar{H}\phi^a t^aH
     \frac{(\bar{H}H)^{n'-1}}{M_{\mathrm{Pl}}^{2n'}}\bar{H}= 0 \ , \\
      -F_{\bar{H}}^\dagger &= -n\lambda_Y \frac{(\bar{H}H)^{n-1}}{M_\mathrm{Pl}^{2n-2}}Y H + \zeta \phi^a t^aH \frac{(\bar{H}H)^{n'}}{M_{\mathrm{Pl}}^{2n'}} +
      n'\zeta \bar{H}\phi^a t^aH
     \frac{(\bar{H}H)^{n'-1}}{M_{\mathrm{Pl}}^{2n'}}H= 0 \ .
\end{align}
Hereafter, we use the same symbols
for the chiral superfields and their scalar components.
For convenience of description, we introduce the following parameters,
\begin{align}
    \hat{v}_X = \frac{v_X}{\lambda_X^{1/2}}\ , \quad 
      \hat{v}_Y = \frac{v_Y}{\lambda_Y^{1/2}}\ .
\end{align}
From the above $F$-term conditions, we find a supersymmetric vacuum at
\begin{gather}
\label{eq:VEVs}
    \expval{\phi^{a}} = \qty(\hat{v}_XM_\mathrm{Pl}^{n-1})^{1/n}\delta^{a3} \ ,\quad 
   \expval{H}
      =\mqty( \qty(\hat{v}_YM_\mathrm{Pl}^{n-1})^{1/n}\\0)\ , 
    \quad 
   \expval{\bar{H}} =  \mqty(
    \qty(\hat{v}_Y M_\mathrm{Pl}^{n-1})^{1/n}\\0)\ , \\
 \expval{X} = \frac{\zeta M_\mathrm{Pl}}{4 n \lambda _{X}}
\qty(\frac{\hat{v}_Y^2}{ M_\mathrm{Pl}^2})^{\frac{(1+n')}{n}} \qty(\frac{\hat{v}_X^2}{M_\mathrm{Pl}^2})^{\frac{(1-2n)}{2n}}
\ ,
     \quad \expval{Y} = \frac{(1+n')\zeta M_\mathrm{Pl}}{2 n \lambda _{Y}}
\qty(\frac{\hat{v}_Y^2}{ M_\mathrm{Pl}^2})^{\frac{(1-n+n')}{n}} \qty(\frac{\hat{v}_X^2}{M_\mathrm{Pl}^2})^{\frac{1}{2n}  
}\ .
\end{gather}
This vacuum also satisfies 
the $D$-term conditions,
\begin{align}
    D^a = -ig \epsilon^{abc}\phi^{b\dagger}\phi^c + 
    gH^\dagger t^a H - g\bar{H} t^a \bar{H}^\dagger = 0 \ .
\end{align}

At the above vacuum, we see that 
SU(2) is broken down to the U(1) subgroup, $\uoneG$, by the vacuum expectation value (VEV) of $\phi^3$.  
The $\uoneG$ is broken by the VEVs of the doublets (see \cref{fig:timeline}).
Accordingly, $v_\mathrm{SU(2)}$ and $v_\mathrm{U(1)}$ are given by,
\begin{align}
\label{eq:breakingscales}
    v_\mathrm{SU(2)}= (\hat{v}_X M_\mathrm{Pl}^{n-1})^{1/n}\ , \quad 
        v_\mathrm{U(1)}= (\hat{v}_Y M_\mathrm{Pl}^{n-1})^{1/n}\  ,
\end{align}
respectively.

Due to the nature of the higher-dimensional interactions, the VEVs of $X$ and $Y$ tend to be  large.
In the following analysis, we expand fields under the assumption that all field values are sufficiently smaller than $M_\mathrm{Pl}$. 
The conditions $|\expval{X}|,\,|\expval{Y}|\ll M_{\mathrm{Pl}}$ are satisfied when $\zeta$ is rather suppressed,
\begin{align}
\label{eq:zeta condition}
    \zeta < \min\qty[\lambda_X \qty(    \frac{v_\mathrm{SU(2)}^{2n}}{v_\mathrm{U(1)}^{2n-2}M_\mathrm{Pl}^2})\ , \,
    \lambda_Y     \qty(\frac{v_\mathrm{U(1)}^3}{v_\mathrm{SU(2)}M_\mathrm{Pl}^2})
    ] = \lambda_{X,Y}\times \order{10^{-3}}  \ ,
\end{align}
with which 
\begin{align}
\label{eq:XandYVEVs}
    X < \frac{1}{4n}\times v_\mathrm{SU(2)}\ , \quad Y < \frac{n-1}{2n} \times v_\mathrm{U(1)}\ .
\end{align}

At this vacuum, 
the discrete $\mathbb{Z}_{4nR}$ symmetry 
is broken down to a $\mathbb{Z}_{2R}$ symmetry.
Consequently, the superpotential acquires a non-zero vacuum expectation value,
\begin{align}
\langle W\rangle = \zeta v_\mathrm{SU(2)}v_\mathrm{U(1)}^{2+2n'}  \sim \zeta\times \frac{(10^{16}\,\mathrm{GeV})^{3+2n'} }{M_\mathrm{Pl}^{2n'}}
    \ .
    \label{eq:WVEV}
\end{align}
In \cref{sec:gravitino}, we will see that $\langle W \rangle$ 
can provide the appropriate 
gravitino mass in the hundreds of TeV to the PeV range.
Finally but not least,
we require that the vacuum energy density is canceled between the contribution 
from the supersymmetry breaking sector and that from the VEV of the superpotential.
This fine-tuning of the cosmological constant is a common issue in conventional supersymmetric models.

\subsection{Mass Spectrum at Vacuum}
Let us discuss the mass spectrum of the inflaton sector. 
In this subsection,
the chiral superfields are shifted by the VEVs \eqref{eq:VEVs} so that they become zeros at the vacuum.
The SU(2) vector multiplet obtain masses, 
\begin{align}
\label{eq:Gauge boson mass}
    M_{a=1,2}^2 = g^2\qty(2 v_\mathrm{SU(2)}^2 +  v_\mathrm{U(1)}^2)\ ,\quad
    M_{a=3}^2 = g^2 v_\mathrm{U(1)}^2\ .
\end{align}
The corresponding would-be Goldstone modes are 
three zero modes of 
the superpotential:
\begin{align}
    G_1 & = -\frac{i}{\sqrt{2+4 r_V^2}}H_2-\frac{i}{\sqrt{2+4
   r_V^2}}\bar{H}_2-i\frac{\sqrt{2}r_V}{\sqrt{1+ 2r_V^2}} \phi_1\ , \\
    G_2 & = \frac{1}{\sqrt{2+4 r_V^2}}H_2-\frac{1}{\sqrt{2+4
   r_V^2}}\bar{H}_2+i\frac{\sqrt{2}r_V}{\sqrt{1+ 2r_V^2}} \phi_2\ ,  \\
   G_3 &= \frac{1}{\sqrt{2}}(H_1 -\bar{H}_1) \ ,
\end{align}
where $r_V = v_\mathrm{SU(2)}/v_\mathrm{U(1)}$. 
In the large $r_V$ limit, we find that the would-be Goldstone modes for 
the breaking $\SU(2)\to\uoneG$ is dominated by $\phi_{1,2}$.

The remaining six massive chiral fields are given by,
\begin{align}
       \chi_1 &=\frac{r_V}{\sqrt{1+2 r_V^2}}H_2+\frac{r_V}{\sqrt{1+2
   r_V^2}}\bar{H}_2-\frac{1}{\sqrt{1+ 2r_V^2}} \phi_1\ , \\
    \chi_2 &=-\frac{ir_V}{\sqrt{1+2 r_V^2}}H_2+\frac{ir_V}{\sqrt{1+2
   r_V^2}}\bar{H}_2-\frac{1}{\sqrt{1+ 2r_V^2}} \phi_2\ ,   \\
    \chi_3 &= \frac{1}{\sqrt{2}}(H_1+\bar{H}_1)\  ,  
\end{align}
and $\phi_3$, $X$ and $Y$.
They obtain supersymmetric masses as, 
\begin{align}
\label{eq:Wmassive}
    W|_\mathrm{mass} =& 
    - \frac{2n\lambda_X v_\mathrm{SU(2)}^{2n-1} }{M_\mathrm{Pl}^{2n-2}}X\phi_3
      - \frac{\sqrt{2}n\lambda_Yv_\mathrm{U(1)}^{2n-1}}{M_\mathrm{Pl}^{2n-2}}Y\chi_3
      + \frac{1+n'}{\sqrt{2}}\frac{\zeta v_\mathrm{U(1)}^{1+2n'}}{M_\mathrm{Pl}^{2n'}}
      \chi_3 \phi_3 -\frac{2n-1}{4r} \frac{\zeta v_\mathrm{U(1)}^{1+2n'}}{M_\mathrm{Pl}^{2n'}} \phi_3^2 
       \cr
      &- \frac{1+2r_V^2}{4r_V} \frac{\zeta v_\mathrm{U(1)}^{1+2n'}}{M_\mathrm{Pl}^{2n'}} \chi_1^2 - \frac{1+2r_V^2}{4r_V} \frac{\zeta v_\mathrm{U(1)}^{1+2n'}}{M_\mathrm{Pl}^{2n'}} \chi_2^2+\frac{1}{2}
      (1+n')(1-n+n')r_V\frac{\zeta v_\mathrm{U(1)}^{1+2n'}}{M_\mathrm{Pl}^{2n'}}  \chi_3^2 
      \ .
\end{align}
Their numerical values will be given in the next section.

Note that $X,Y,\phi_3, \chi_{1,2,3}$
and $\Im[G_{1,2,3}]$ 
do not obtain masses from the $D$-term potential.
On the other hand,
$\Re[G_{1,2}]$ 
and $\Re[G_3]$ 
obtain masses $M_{a=1,2}$ and $M_{a=3}$ in \cref{eq:Gauge boson mass} 
from the $D$-term potential.
This means that, only $\Im[G_{1,2,3}]$ are the would-be Goldstone modes in the component field formalism, while chiral $G_{1,2,3}$ supermultiplets form the would-be goldstone multiplets in the superfield formalism.

\subsection{Remaining U(1) Symmetry Breaking after Inflation}
\label{sec:U1breaking}
During inflation, $H$ and $\bar{H}$ stay at the origin by assuming the positive Hubble mass terms, as we discuss later (see \cref{eq:mHeff}).
After inflation, on the other hand, 
they roll down to the vacuum \eqref{eq:VEVs} by the negative Hubble mass terms.

Notice that the superpotential
term proportional to $\zeta$ 
in \cref{eq:Wpot}
leads to a scalar potential
of $H=\bar{H}$,
\begin{align}
   V_{\zeta} 
   \sim 
   \frac{
   \zeta^2 \expval{\phi_3}^2}
   {M_\mathrm{Pl}^{4n-8}}
  \qty| H^{4n-6}|\ .
\end{align}
In combination with the negative Hubble mass term, this term may form a local minimum other than the supersymmetric vacuum in \cref{eq:VEVs}.
To avoid this, let us compare $V_\zeta$ with the destabilization terms of $H$ and $\bar{H}$ in the scalar potential
\begin{align}
    V_{\mathrm{destabilize}} = -\lambda_Y \frac{v_Y^2(\bar{H}H)^{n}}{M_{\mathrm{Pl}}^{2n-2}} + \hc \ ,
\end{align}
which drives $H$ and $\bar{H}$ to the supersymmetric vacuum.
For $n>3$,
$V_\zeta$ is more suppressed 
than $V_\mathrm{destabilize}$,
and hence, $V_\zeta$ does not 
lead to a local minimum.
For $n=3$, on the other hand, we require 
\begin{align}
\label{eq:smooth}
    \zeta^2 \frac{v_\mathrm{SU(2)}^2}{M_\mathrm{Pl}^4} \lesssim \lambda_Y \frac{v_Y^2}{M_\mathrm{Pl}^{4}}\ ,
\end{align}
so that $V_\zeta$ is subdominant. 
In fact, this condition 
is satisfied under the condition in \cref{eq:zeta condition}.
As a result, we find that 
$H$ and $\bar{H}$ smoothly roll down to the supersymmetric vacuum after inflation for $n\ge 3$.%
\footnote{For $n=2$, we may take $n'=n+2$, with which smooth U(1) symmetry breaking is achieved.}

\section{Inflation Dynamics}
\label{sec:inflation}
In the previous section,
we have discussed the vacuum structure and the mass spectrum.
In this section, we discuss 
the dynamics of the new inflation.

\begin{figure}[t]
    \centering
\includegraphics[width=0.45\linewidth]{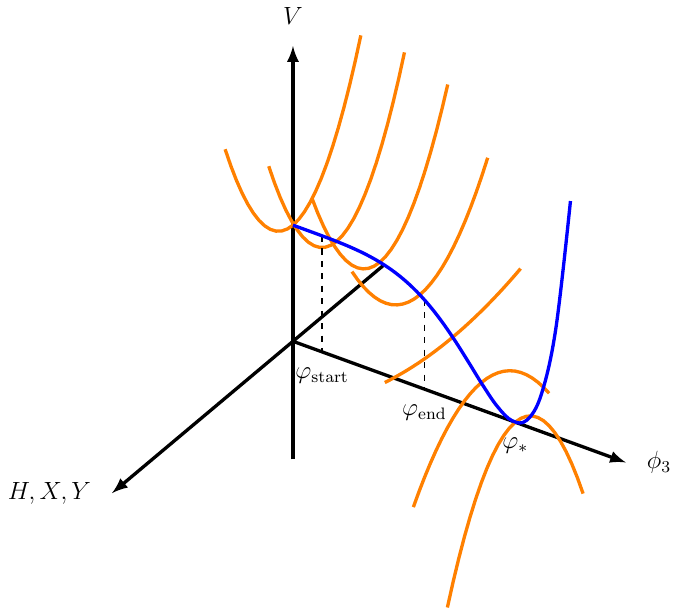}
    \caption{A schematic picture of the scalar potential of the inflaton sector.
    The new inflation starts from a small field value $\varphi_{\mathrm{start}}$ and ends at $\varphi_\mathrm{end}$ where the slow-roll condition is violated.
    \modifiedat{JCAP2}{Unlike hybrid inflation, the other scalar fields remain stabilized at the origin and roll down the potential only after inflation ends. These fields acquire masses larger than the Hubble scale during inflation (see Sec.\,\ref{sec:Stabilization}), making the model effectively single-field inflation.
    }}
    \label{fig:potential}
\end{figure}

In \cref{fig:potential},
we show the inflaton sector potential schematically.
The field $\phi_3$ plays the role of the inflaton of the new inflation starting from a small field value.
We also assume that $H$, $\bar{H}$, 
$X$ and $Y$ are stabilized at around the origin by the Hubble induced mass terms during inflation.
Unlike in the hybrid inflation, 
$H$, $\bar{H}$, $X$ and $Y$ start to roll down the potential well after the end of inflation.

\subsection{Inflaton Potential}
The K\"ahler potential terms relevant for the inflation dynamics are as follows:
\begin{align}
\label{eq:Kpot}
    K =& \phi^\dagger \kdot \phi + H^\dagger\kdot  H  + \bar{H}^\dagger\kdot  \bar{H}  
    + X^\dagger X + Y^\dagger Y \cr
    & +
    \frac{\kappa_X}{4}
    \frac{(X^\dagger X)^2}{M_\mathrm{Pl}^2}      + 
    \frac{\kappa_Y}{4}
    \frac{(Y^\dagger Y)^2}{M_\mathrm{Pl}^2}
    + 
    \kappa_{XY}
    \frac{(Y^\dagger Y)(X^\dagger X)
    }{M_\mathrm{Pl}^2}
    \cr
&     +
    k_\phi
    \frac{(X^\dagger X)(\phi^\dagger \kdot \phi)}{M_\mathrm{Pl}^2}      + 
    k_H
    \frac{(X^\dagger X)(H^\dagger \kdot H)}{M_\mathrm{Pl}^2}      + 
     k_{\bar{H}}
    \frac{(X^\dagger X)(\bar{H}^\dagger \kdot \bar{H})}{M_\mathrm{Pl}^2}   \cr
&      +
    f_\phi
    \frac{(Y^\dagger Y)(\phi^\dagger \kdot \phi)}{M_\mathrm{Pl}^2}      + 
    f_H
    \frac{(Y^\dagger Y)(H^\dagger \kdot H)}{M_\mathrm{Pl}^2}      + 
     f_{\bar{H}}
    \frac{(Y^\dagger Y)(\bar{H}^\dagger \kdot \bar{H})}{M_\mathrm{Pl}^2} 
    \ .
\end{align}
Here, ``$\kdot$" implicitly includes $e^{-2V}$, where $V$ is the SU(2) vector superfield.
The parameters $\kappa$'s,  $k$'s and $f$'s are $\order{1}$ coefficients.
In \cref{eq:Kpot}, we omitted several possible terms where $X$ and $Y$ are interchanged, as their inclusion would not lead to essential differences. Additionally, we have also omitted the four-point interactions of SU(2) charged fields, as they do not affect the inflation dynamics.

By taking the supergravity effects into account, the superpotential in \cref{eq:Wpot} and the K\"ahler potential in \cref{eq:Kpot} leads 
to the dominant part of the inflaton potential,
\begin{align}
\label{eq:Vphi}
    V_\phi \simeq& 
    \qty|\qty(v_X^2-\lambda_X
    \frac{(\phi\cdot\phi)^n}{M_{\mathrm{Pl}}^{2n-2}})|^2
    \times \qty(1-(k_\phi-1)\frac{\phi^\dagger\cdot\phi}{M_\mathrm{Pl}^2})\cr
    &+  v_Y^4  \times \qty(1-\frac{f_\phi-1}{M_\mathrm{Pl}^2}\phi^\dagger\cdot \phi) 
    + \cdots\ ,
\end{align}
for $|\phi|\ll M_\mathrm{Pl}$.
Here, we have set all the fields other than $\phi^a$ to zero, which will be justified later.
The terms in the first and the second lines come from the $F$-components of $X$ and $Y$, respectively.
We omitted the soft supersymmetry breaking masses of $\order{m_{3/2}}$, since they are much smaller than the Hubble scale during inflation.

As we will see shortly, the inflation ends  
at $|\phi|\ll v_{\mathrm{SU(2)}}$.
Thus, the Hubble parameter during inflation is given by,
\begin{align}
    H^2_\mathrm{inf}\simeq
    \frac{1}{3M_\mathrm{Pl}^2}\qty(v_X^4 +
    v_Y^4)\ .
\end{align}
Parametrically, we assume
\begin{align}
\label{eq:dominance}
    v_X^4 \gg v_Y^4\ ,
\end{align}
so that the contribution from the $F$-component of $X$ dominates the inflaton energy density.%
\footnote{We have confirmed that the
$v_Y^4$ contribution to the energy density during inflation does not affect the following discussion for $v_Y^4/v_X^4\lesssim 0.05$.}

For a small $|\phi|$, 
the inflaton mass term is given by,
\begin{align}
    V_\phi|_\mathrm{mass} \simeq -m_{\phi,\,\mathrm{eff}}^2\,\phi^\dagger \cdot \phi
    =-\qty(\frac{(k_\phi-1)v_X^4}{M_\mathrm{Pl}^2} +
    \frac{(f_\phi-1)v_Y^4}{M_\mathrm{Pl}^2}
    ) \phi^\dagger \cdot \phi\ ,
\end{align}
where $m_{\phi,\,\mathrm{eff}}$ is typically  $m_{\phi,\,\mathrm{eff}}= \order{H_{\mathrm{inf}}}$.
Thus, for a successful slow-roll inflation, we must require that the parameters are tuned
so that
\begin{align}
\eta_0 :=   \frac{m_{\phi,\,\mathrm{eff}}^2}{3H_\mathrm{inf}^2} \simeq -(k_\phi-1) = \order{10^{-2}}\ .
\end{align}
This tuning is called the $\eta$-problem, although 
the level of fine-tuning is not very severe.

\subsection{Slow-Roll Inflation}
Let us now take the inflaton direction as
\begin{align}
    \phi = (0,0,\varphi/\sqrt{2})\ ,\quad \varphi \in \mathbb{R}\ .
\end{align}
In \cref{eq:Vphi}, the relevant part of the inflaton potential is given by
\begin{align}
\label{eq:Vvarphi}
    V_\varphi =  v_X^4 + \frac{1}{2} \eta_0 \frac{v_X^4}{M_\mathrm{Pl}^2} \varphi^2
    - \frac{1}{2^{n-1}} \frac{\lambda_X v_X^2}{M_\mathrm{Pl}^{2n-2}} \varphi^{2n} + \cdots\ ,
\end{align}
where the other terms are negligibly small. We have also omitted the contribution from $v_Y^4$.
The dynamics of this type of inflaton potential have been analyzed in Ref.\,\cite{Ibe:2006fs,Harigaya:2013pla} (see also Ref.\,\cite{Izawa:1996dv} for earlier work).%
\footnote{In these references, the parameter $n$ corresponds to $2n$ in the present model. The parameters $v^2$ and $g$ in the references also correspond to $v^2$ and $\lambda_X$, respectively.}

The first and the second slow-roll parameters are given by
\begin{align}
    \epsilon_V &:= \frac{M_\mathrm{Pl}^2}{2}\qty(\frac{\partial{V_\phi}/{\partial\varphi}}{V_\phi})^2
    \simeq\frac{1}{2}\qty( \frac{\eta_0\varphi}{M_\mathrm{Pl}}-\frac{n\varphi^{2n-1}}{2^{n-2}\hat{v}_X^2M_\mathrm{Pl}^{2n-3}})^2 
    \ ,\\
    \eta_V&:= M_\mathrm{Pl}^2\frac{\partial^2 V_\phi/\partial \varphi^2}{V_\phi} \simeq \eta_0 - \frac{n(2n-1)\varphi^{2n-2}}{2^{n-2}\hat{v}_X^2M_\mathrm{Pl}^{2n-4}} \ .
\end{align}
For $\eta_0 = \order{10^{-2}}$, the inflation ends when
the inflaton field reaches
\begin{align}
    \varphi_\mathrm{end} \simeq \qty(\frac{2^{n-2} \hat{v}_X^2 M_{\mathrm{Pl}}^{2n-4}}{
   n(2n-1)})^{\frac{1}{2n-2}}\ ,
\end{align}
at which $\eta\simeq -1$.

The relation between the $e$-folding number and the inflaton field value is given by
\begin{align}
    N_e(\varphi) = \int_{\varphi_\mathrm{end}}^\varphi  \frac{d\varphi}{\partial V/\partial\varphi}
    =\frac{1}{-2(n-1)\eta_0}
    \log
    \qty[
    \frac{n-\eta_0 2^{n-2} \hat{v}_X^2 \varphi^{2-2 n} M_{\mathrm{Pl}}^{2 n-4}}{n-\eta_0 2^{n-2} \hat{v}_X^2 \varphi_\mathrm{end}^{2-2 n} M_{\mathrm{Pl}}^{2 n-4}}]\ ,
\end{align}
or
\begin{align}
   \varphi(N_e)  = \varphi_\mathrm{end}
  \times e^{\eta_0 N_e} \qty(\frac{-\eta_0 (2 n-1)}{1-e^{2 \eta_0 (n-1) N_e}-\eta_0 (2
   n-1)})^{\frac{1}{2n-2}} \ .
\end{align}
For given $N_e$, the amplitude $A_s$ and the spectral index $n_s$ of the curvature perturbation are given by
\begin{align}
\label{eq:As}
    A_s &= \frac{1}{8\pi^2}\frac{H_\mathrm{inf}^2}{\epsilon_V M_\mathrm{Pl}^2 } \simeq \frac{ 2^{\frac{1}{n-1}} (n (2 n-1))^{\frac{3}{n-1}+2} \varphi_{N_e}^2
     }{24 \pi ^2
   \left(n^{\frac{n}{n-1}} (2 n-1)^{\frac{1}{n-1}} \hat{\varphi}_{N_e}^{2 n}-\varphi_{N_e}^2
   \eta_0 (n (2 n-1))^{\frac{n}{n-1}}\right)^2} \times \lambda_{X}^2 \qty(\frac{\hat{v}_X}{M_{\mathrm{Pl}}})^{4-\frac{2}{n-1}}\ ,  \\
   n_s &= 1-6\epsilon_V+2\eta_V\simeq 1+2\eta_0 - 2 \hat{\varphi}_{N_e}^{2n-2}\ ,
\end{align}
respectively.
Here, we have defined $\hat{\varphi}_{N_e}:=\varphi(N_e)/\varphi_\mathrm{end}$.
Thus, for a given value of $A_s$, the parameter $\hat{v}_X$ is determined as
\begin{align}
  \hat{v}_X \propto
  \qty(\frac{A_s}{\lambda_X^2})^{\frac{n-1}{4 n-6}}\times M_\mathrm{Pl}\ 
 .
\end{align}

\begin{figure}[t]
    \centering
    \begin{minipage}{0.48\linewidth}  
        \centering        \includegraphics[width=\linewidth]{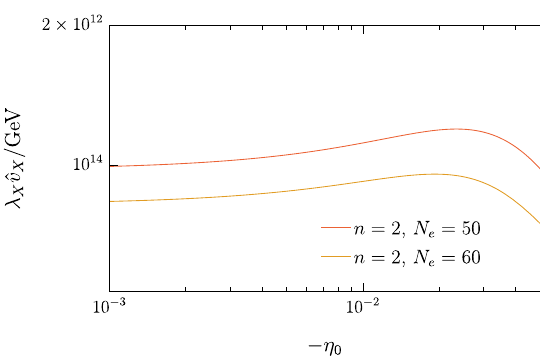}
    \end{minipage}
        \begin{minipage}{0.48\linewidth}          \centering
        \includegraphics[width=\linewidth]{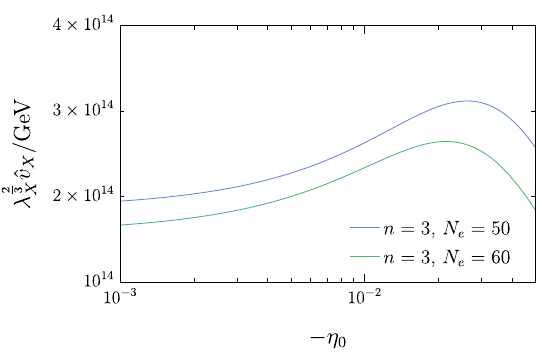}
    \end{minipage}
       \caption{ The required value of $\hat{v}_X$ to reproduce the observed value of the curvature perturbation, $A_s = (2.105 \pm 0.030)\times 10^{-9}$  for $n=2$ (left) and $n=3$ (right).
       }
    \label{fig:scale}
\end{figure}

\Cref{fig:scale} shows the value of $ \lambda_X^{\frac{n-1}{2n-3}}\hat{v}_\mathrm{X}$ 
which reproduces the observed value of the curvature perturbation~\cite{Planck:2018vyg},
\begin{align}
\label{eq:AsVAL}
    A_s =  (2.105 \pm 0.030)\times 10^{-9}\ , \quad (1\sigma\ \mathrm{range})\ .
\end{align}
This value is evaluated at the pivot scale $k = 0.05\,\mathrm{Mpc}^{-1}$.
In the figure, $N_e$ denotes 
the total $e$-folding number,%
\footnote{To estimate $A_s$ in \cref{eq:AsVAL}, we take $\varphi_{Ne(k)}$ with $N_e(k) = N_e-\log(k/H_0)$ where $k=0.05\,\mathrm{Mpc}^{-1}$.}
\begin{align}
\label{eq:Ne}
    N_e \simeq 51+
    \frac{1}{3}\log\qty(\frac{T_R}{10^4\,\mathrm{GeV}})+\frac{1}{3}\log\qty(\frac
    {H_\mathrm{inf}}{10^{10}\,\mathrm{GeV}})\ .
\end{align}
The reheating temperature $T_R$ is related to the Hubble parameter just after reheating by
\begin{align}
    H_R = \qty(\frac{\pi^2g_*(T_R)}{90})^{1/2}\frac{T_R^2}{M_\mathrm{Pl}}\ ,
\end{align}
where $g_*(T_R)\simeq \order{10^2}$ represents the effective degrees of freedom at $T_R$.
In this paper,  we adopt the effective degrees of freedom of the Standard Model (SM) particles as provided in Ref.\,\cite{Saikawa:2018rcs,*Saikawa:2020swg}.

The figure shows that the observed curvature perturbation is achieved for 
\begin{equation}
    \begin{cases}
    \lambda_X \hat{v}_X \simeq  10^{12}\,\mathrm{GeV}& (n=2) \\
\label{eq:vXhatn3}
    \lambda_X^{\frac{2}{3}} \hat{v}_X \simeq(2\mathchar`-3)\times 10^{14}\,\mathrm{GeV}& (n=3)\ .
    \end{cases}
\end{equation}
Correspondingly, the SU(2) breaking scale is estimated as
\begin{align}
   \begin{cases}
    v_\mathrm{SU(2)} \simeq \lambda_X^{-1/2}\times 10^{15}\,\mathrm{GeV}& (n=2)  \\
    v_\mathrm{SU(2)} \simeq \lambda_X^{-2/9}\times 10^{17}\,\mathrm{GeV}&(n=3) \ .
    \end{cases}
\end{align}
Therefore, the SU(2) breaking scale can be close to the conventional GUT scale to reproduce the observed curvature perturbation.

\begin{figure}[t]
    \centering
    \begin{minipage}{0.48\linewidth}  
        \centering        \includegraphics[width=\linewidth]{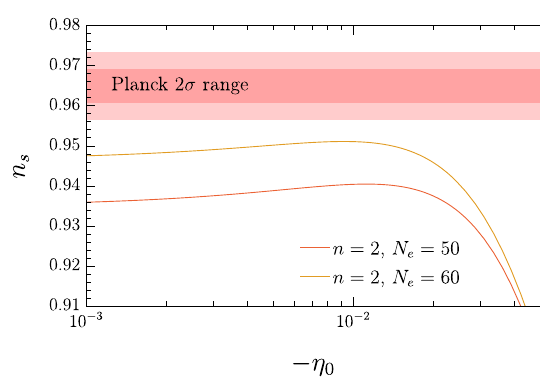}
    \end{minipage}
        \begin{minipage}{0.48\linewidth}          \centering
        \includegraphics[width=\linewidth]{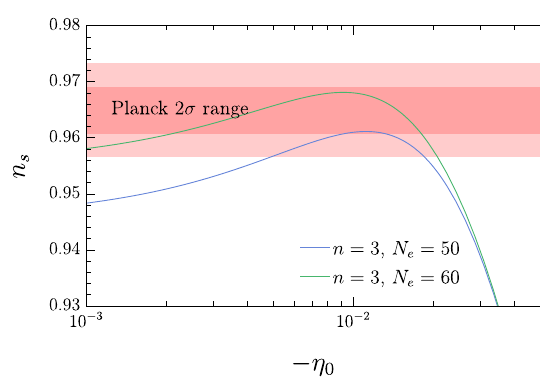}
    \end{minipage}
       \caption{
    The spectral index of the curvature perturbation at the pivot scale $k_0=0.05\,\mathrm{Mpc}^{-1}$. 
       The pink shaded region shows the $2\sigma$ range of $n_s$ from the Planck result\,\cite{Planck:2018vyg}.
       }
    \label{fig:ns}
\end{figure}

In \cref{fig:ns}, we also show the spectral index
at the pivot scale 
$k = 0.05$\,Mpc$^{-1}$.
The figure shows that the spectral index 
is smaller than the 2$\sigma$ range of the observed value~\cite{Planck:2018vyg},
\begin{align}
  n_s=  0.9665\pm0.0038\ ,\quad (68\%\,\mathrm{limits})\ ,
\end{align}
for $n=2$.
For $n=3$, on the other hand, the spectral index can be consistent with the observation for $\eta_0 \simeq - 0.01$ and $N_e \simeq 50$--$60$.
Thus, 
in the following, we focus on the model with $n=3$.%
\footnote{As pointed out in Refs.\,\cite{Nakayama:2011ri,Nakayama:2012dw}, when the scale of soft supersymmetry breaking is close to $H_\mathrm{inf}$, 
a similar model with $n=2$ 
can achieve a spectral index consistent with observational values
due to radiative corrections 
to the scalar potential.
}

\subsection{Stabilization of Fields  Other than Inflaton}
\label{sec:Stabilization}
In the above discussion, we have assumed that only the inflaton has a non-vanishing field value during inflation. Now, let us discuss \modifiedat{JCAP2}{how the} other fields are stabilized at around $X=Y=H=\bar{H}=0$.
The potential of $H$ and $\bar{H}$ is given by
\begin{align}
    V_{H,\bar{H}}=&
    \qty|
    v_X^2-\lambda_X
    \frac{(\phi\cdot\phi)^n}{M_{\mathrm{Pl}}^{2n-2}}|^2
    \times \qty((1-k_H)\frac{H^\dagger H}{M_\mathrm{Pl}^2} + (1-k_{\bar{H}})\frac{\bar{H}^\dagger \bar{H}}{M_\mathrm{Pl}^2})
    \cr
    &+\qty|
    v_Y^2-\lambda_Y\frac{(\bar{H}H)^n}{M_\mathrm{Pl}^{2n-2}}|^2\times
    \qty(1+(1-f_H)\frac{H^\dagger H}{M_\mathrm{Pl}^2}+(1-f_{\bar{H}})\frac{\bar{H}^\dagger \bar{H}}{M_\mathrm{Pl}^2})
    -v_Y^4 \cr
    & +\qty(\qty|
    v_Y^2-\lambda_Y\frac{(\bar{H}H)^n}{M_\mathrm{Pl}^{2n-2}}|^2-v_Y^4)\times (1-f_\phi)\frac{\phi^\dagger \cdot \phi}{M_\mathrm{Pl}^2}\ .
\end{align}
Here, we have neglected the contribution from the superpotential term proportional to $\zeta$ since they are suppressed.
We have also omitted the soft
supersymmetry breaking masses of $H$ and $\bar{H}$, since the Hubble rate $H_\mathrm{inf}$
is much larger.

The above potential leads to the effective mass terms around $H=\bar{H}=0$, 
\begin{align}
\label{eq:mHeff}
  m_{H,\bar{H}}^2 =\frac{1}{M_\mathrm{Pl}^2}  \qty|
  v_X^2-\lambda_X
    \frac{(\phi\cdot\phi)^n}{M_{\mathrm{Pl}}^{2n-2}}|^2
    (1-k_{H,\bar{H}}) + \frac{1}{M_\mathrm{Pl}^2}v_Y^4(1-f_{H,H})\ .  
\end{align}
Since we have assumed $v_X^4\gg v_Y^4$,
the first term provides the dominant 
contribution to the effective mass during inflation.
To stabilize $H$ and $\bar{H}$, we assume
\begin{align}
1-k_{H,\bar{H}}> \frac{3}{4}\ ,
\end{align}
where the lower limit is given to suppress quantum fluctuations
in the de-Sitter vacuum~\cite{Birrell:1982ix}.

For successful $\uoneG$ symmetry breaking after inflation,
on the other hand,
we also require that the second term in \cref{eq:mHeff} gives the negative mass squared to $H$ and $\bar{H}$,
\begin{align}
   1- f_{H,\bar{H}} < 0 \ .
\end{align}
This negative mass squared dominates the Hubble induced mass well after inflation.
Then, given the condition in \cref{eq:smooth}
is satisfied,
$H$ and $\bar{H}$
smoothly roll down to the supersymmetric vacuum.

Next, let us consider the potential of $X$ and $Y$. 
Along the inflaton trajectory, the Hubble induced masses of $X$ and $Y$ are given by,
\begin{align}
\label{eq:VXY}
m_X^2 &= 
\frac{1}{M_\mathrm{Pl}^2}\qty|
  v_X^2-\lambda_X
    \frac{(\phi\cdot\phi)^n}{M_{\mathrm{Pl}}^{2n-2}}|^2(-\kappa_X)+ \frac{1}{M_\mathrm{Pl}^2}v_Y^4  (1-\kappa_{XY}) \ , \\
m_Y^2 & =   
\frac{1}{M_\mathrm{Pl}^2}\qty|
  v_X^2-\lambda_X
    \frac{(\phi\cdot\phi)^n}{M_{\mathrm{Pl}}^{2n-2}}|^2 (1-\kappa_{XY}) + 
\frac{1}{M_\mathrm{Pl}^2}v_Y^4(-\kappa_Y)   \ .
\end{align}
Thus, we require
\begin{align}
\label{eq:kappa conditions}
-\kappa_{X} > \frac{3}{4}\ ,
\quad 1-\kappa_{XY} >\frac{3}{4}\ ,
\end{align}
to stabilize $X$ and $Y$ during inflation as in the case of $H$ and $\bar{H}$.

The Hubble induced mass proportional to $v_X^4$ 
in 
\cref{eq:VXY} vanishes when $\phi_3$ settles to its minimum after inflation.
Then, the $v_Y^4$ contribution in 
$m_X^2$ remains positive even after inflation due to the condition in \cref{eq:kappa conditions}.
However, $X$ also obtains 
supersymmetric mass around 
$\expval{X}$ at $\phi_3 = v_\mathrm{SU(2)}$,
\begin{align}
\label{eq:MX}
    M_{X\phi_3} = \frac{2n\lambda_X v_\mathrm{SU(2)}^{2n-1} }{M_\mathrm{Pl}^{2n-2}}\ ,
\end{align}
which stems from the $|F_{\phi_3}|^2$
contribution
(i.e. the first term in \cref{eq:Wmassive}).
With $|M_{X\phi_3}|^2 \gg m_X^2$,
$X$ smoothly rolls down to the vacuum even if $m_X^2$ is positive.
We will confirm shortly that $|M_{X\phi_3}|^2\gg |m_X|^2$ for reference scenarios.
Similarly, the Hubble
induced mass of $Y$ does not affect the dynamics after inflation.

\subsection{Reference Values in  \texorpdfstring{$\boldsymbol{n=3}$}{n=3} Model}
\label{sec:reference}
As discussed above, the model with $n=3$ can simultaneously reproduce the amplitude and spectral index of the observed curvature perturbation for
$\eta_0 \simeq -10^{-2}$
and $\hat{v}_X$ in \cref{eq:vXhatn3}.
In this subsection, we provide a summary of the reference parameter values for the subsequent analysis.
As a reference value, we take
\begin{align}
    \lambda_X^{2/3} \hat{v}_X = 3\times 10^{14}\,\mathrm{GeV}\ ,
\end{align}
(see Figs.\,\ref{fig:scale} and \ref{fig:ns}).
The corresponding SU(2) breaking scale is
\begin{align}
    v_\mathrm{SU(2)} = (\hat{v}_X M_\mathrm{Pl}^2)^{1/3} = \lambda_{X}^{-2/9}\times 1.2\times 10^{17}\,\mathrm{GeV} \ .
\end{align}
Note that, as mentioned earlier, 
if the cutoff scale is below $M_\mathrm{Pl}$, $\lambda_X$ need not be of $\order{1}$ (see \cref{sec:largelam}).
Here, we take $\lambda_X=100$ as a reference, which yields
\begin{align}
\label{eq:SU2reference}
    v_\mathrm{SU(2)} = 4.4\times 10^{16}\,\mathrm{GeV}
    \times
    \qty(\frac{\lambda_{X}}{100})^{-2/9}
    \ .
\end{align}
Then, by using the PTA favored parameters,
the VEV for $\uoneG$ symmetry breaking is given by
\begin{align}
\label{eq:U1reference}
    v_\mathrm{U(1)} = 7.0\times 10^{16}\,\mathrm{GeV}\times 
r_\kappa^{-1}\qty(\frac{f_{T}}{0.15})^{-1/2}
      \qty(\frac{\lambda_{X}}{100})^{-2/9}\ ,
\end{align}
for $\sqrt{\kappa}\simeq 8$ (see \cref{eq:rootkappa}).
Note that we have used the BPS limit of the monopole mass, i.e., $f_M = 1$, in deriving \cref{eq:U1reference}, as the scalar potential of the triplet is relatively flat.
The nominal value of $v_{\mathrm{U}(1)}$ in \cref{eq:U1reference} corresponds to $G_{\mathrm{N}}\mu_\mathrm{str}\simeq 10^{-4.5}$,
consistent with the 
preferred range of $G_\mathrm{N}\mu_\mathrm{str}$
from the PTA signal in \cref{eq:Gmu range}.
Hereafter, to simplify notation, we introduce
\begin{align}
\lambda_{X100}= \frac{\lambda_X}{100} \ ,
\quad
    f_{T0.15} = \frac{f_T}{0.15}\ .
\end{align}

By substituting $v_{\mathrm{U(1)}}$
and $v_{\SU(2)}$ into \cref{eq:breakingscales}, we find 
\begin{align}
    \hat{v}_Y &= 5.9\times 10^{13}\,\mathrm{GeV} 
   \times r_\kappa^{-3}  f_{T0.15}^{-3/2} \times  \lambda_{X100}^{-2/3}\ .
\end{align}
and hence, 
\begin{align}
\label{eq:vXref}
    v_X &= \lambda_X^{1/2} \hat{v}_X = 1.4 \times 10^{14}\, \mathrm{GeV}\times \lambda_{X100}^{-1/6}\ , \\
    v_Y&=\lambda_Y^{1/2} \hat{v}_Y =   5.9\times 10^{13}\,\mathrm{GeV}
     \times \lambda_Y^{1/2}
        \times r_\kappa^{-3}
  f_{T0.15}^{-3/2}\times \lambda_{X100}^{-2/3} \ .
\end{align}
Thus, we can easily achieve the condition $|v_X|^4 \gg |v_Y|^4$ during inflation for $\lambda_Y r_\kappa^{-6}\lesssim \order{10^{-1}}$.
The corresponding Hubble scale during inflation is 
\begin{align}
\label{eq:HubbleRef}
    H_\mathrm{inf} \simeq \frac{v_X^2}{\sqrt{3}M_\mathrm{Pl}} \simeq 
    4.7\times 10^{9}\, \mathrm{GeV}\times \lambda_{X100}^{-1/3}\ .
\end{align}

Since $\lambda_X\gg \lambda_Y$, the upper limit \eqref{eq:zeta condition} on $\zeta$ reduces to
\begin{align}
\label{eq:zetamax}
    \zeta < \zeta_\mathrm{max}:=1.4\times 10^{-3}\times \lambda_Y\times r_\kappa^{-3} f_{T0.15}^{-3/2}\times  \lambda_{X100}^{-4/9}\ .
\end{align}
By defining $\hat{\zeta}=\zeta/\zeta_\mathrm{max}$, 
the VEV of the superpotential, 
and hence the gravitino mass is given by,
\begin{align}
\label{eq:m32 for n=3}
m_{3/2}=\frac{1}{M_\mathrm{Pl}^2}   \langle W \rangle = 2.1\times 10^{7}\,\mathrm{GeV}\times \lambda_Y\times \hat{\zeta}\times r_\kappa^{-7}
f_{T0.15}^{-7/2}\times  \lambda_{X100}^{-14/9} \ .
\end{align}
Thus, for $\lambda_Y\times \hat{\zeta}\times r_\kappa^{-7}= \order{10^{-(\numrange{1}{2})}}$, 
we can achieve the gravitino mass of $\order{\numrange{0.1}{1}}$\,PeV (see \cref{sec:MSSM}).

The masses in \cref{eq:Wmassive} are  estimated as
\begin{align}
M_{X\phi_3} &=\frac{6\lambda_X v_\mathrm{SU(2)}^{5} }{M_\mathrm{Pl}^{4}} = 2.7\times 10^{12}\,\mathrm{GeV}\times  \lambda_{X100}^{-1/9}\ , \\
M_{Y\chi_3} &= \frac{3\sqrt{2}\lambda_Yv_\mathrm{U(1)}^{5}}{M_\mathrm{Pl}^{4}} = 2.1\times 10^{11}\,\mathrm{GeV}\times \lambda_Y\times r_\kappa^{-5}f_{T0.15}^{-5/2}\times \lambda_{X100}^{-10/9}\ , \\
\label{eq:Mchi12}
M_{\chi_{1,2}}&= \frac{1+2r_V^2}{4r}\frac{\zeta v_\mathrm{U(1)}^{3}}{M_\mathrm{Pl}^{2}} =
6.5 \times 10^{10}\,\mathrm{GeV} \times \lambda_Y\times
\hat{\zeta}\times r_\kappa^{-7} f_{T0.15}^{-7/2}(1+0.76r_\kappa^2f_{T0.15})\times \lambda_{X100}^{-10/9}\ .
\end{align}
The mass terms proportional to $\zeta$ in \cref{eq:Wmassive} are relevant only for $\chi_{1,2}$.
By comparing $M_{X\phi_3}$ and the Hubble rate $H_\mathrm{inf}$, we find that 
the Hubble induced mass is subdominant for $X$ and $Y$ after inflation.
From Eqs.\,\eqref{eq:Mchi12} and \eqref{eq:m32 for n=3}, we also find that there is a relation between $M_{\chi_{1,2}}$ and $m_{3/2}$,
\begin{align}
    M_{\chi_{1,2}} = 3.1\times 10^3\times \lambda_{X100}^{4/9}\times (1+ 0.76 r_\kappa^2 f_{T0.15})\times m_{3/2} \ .
\end{align}

In the following, we define two benchmark points,
\begin{align}
\label{eq:benchmark}
\mathrm{BP}_1 &:
& \lambda_X &= 100\ ,  
&\lambda_Y&=\order{10^{-1}}\ ,  
&\hat{\zeta} &= \order{10^{-1}}   \ , 
&r_\kappa &= 1\ , 
&f_T&=0.15 \\
\mathrm{BP}_2 &: 
&\lambda_X &= 100\ ,  
&\lambda_Y &= \order{10^{2}}\ ,
&\hat{\zeta} &= \order{10^{-1}}  \ , 
&r_\kappa &= 3\ ,  
&f_T&=0.15\ .  
\end{align}
The corresponding string tensions are given by,
\begin{align}
    &\mathrm{BP}_1 :\quad G_\mathrm{N}\mu_\mathrm{str}\simeq 10^{-4.5} \ ,\\
     &\mathrm{BP}_2 :\quad G_\mathrm{N}
\mu_\mathrm{str}\simeq 10^{-5.5} \ . 
\end{align}
In both the benchmark points, the mass spectrum in the inflaton sector satisfies
\begin{align}
\label{eq:spectrum}
    M_{X\phi_3}\ , \ M_{Y\chi_3}   > 2 \times M_{\chi_{1,2}}\ ,
\end{align}
which will be relevant to the reheating process.

\subsection{Breaking of Discrete Symmetries}
\label{sec:domain wall}
As shown in Table\,\ref{tab:symmetry}, the model features two discrete symmetries
$\mathbb{Z}_{4nR}$ and $\mathbb{Z}_{2n}$
that ensure the form of the superpotential in \cref{eq:Wpot}. 
Let us discuss their spontaneous breaking.  

First, we examine the fate of the $\mathbb{Z}_{2n}$ symmetry (e.g., $\mathbb{Z}_6$ for $n=3$). This is spontaneously broken down to a $\mathbb{Z}_2$ symmetry at the onset of inflation by $\phi_3 \neq 0$. 
The doublets, $H$ and $\bar{H}$, are charged under this residual $\mathbb{Z}_2$ symmetry. 
Note that the VEVs of $H$ and $\bar{H}$ do not cause the domain-wall problem, since 
$\mathbb{Z}_2$ is embedded in the $\uoneG$ gauge symmetry and there is no potential barrier between the vacua connected by the $\mathbb{Z}_2$ symmetry.

Next, we consider the $\mathbb{Z}_{4nR}$ symmetry (e.g., $\mathbb{Z}_{12R}$ for $n=3$). 
This symmetry is also spontaneously broken at the onset of inflation by $\phi_3 \neq 0$, apparently down to $\mathbb{Z}_{2R}$. 
However, a larger subgroup remains unbroken along the $\phi_3$ trajectory, consisting of the $\mathbb{Z}_{4R}$ subgroup of $\mathbb{Z}_{4nR}$ and $\sigma_1 \in\SU(2)$.
Under this SU(2) transformation we have:
\begin{align}
\mqty(0\\ 0\\ \phi_3) \to \mqty(0\\0\\-\phi_3) \ , \quad
\mqty(H_1\\H_2) \to \mqty(H_2\\H_1)\ , \quad
\mqty(\bar{H}_1\\ \bar{H}_2) \to \mqty(\bar{H}_2\\ \bar{H}_1)\ .
\end{align}
Combined with $\mathbb{Z}_{4nR}$,
it leads to a residual $\mathbb{Z}_{4R}$ symmetry,
\begin{align}
\phi_3 \to \phi_3\ , \quad X \to -X\ , \quad Y \to -Y\ , \quad H_{1,2} \to -i H_{2,1}\ , \quad \bar{H}_{1,2} \to  i \bar{H}_{2,1}\ .
\end{align}
Therefore, the VEVs of $X$ and $Y$ spontaneously break the $\mathbb{Z}_{4R}$ down to $\mathbb{Z}_{2R}$.
Note that unlike in the case with $\mathbb{Z}_{2n}$ symmetry discussed above, there is a potential barrier between these field points.

This analysis suggests that the breaking of the discrete $R$-symmetry could lead to a domain wall problem.
However, this problem can be evaded if there is another sector 
which provides 
subdominant contributions to the VEV of $W$ 
by spontaneous 
$R$-symmetry breaking.
In fact,
a dynamical supersymmetry breaking sector generically breaks $R$-symmetry spontaneously~\cite{Nelson:1993nf,Bagger:1994hh} (see also Refs.\,\cite{Komargodski:2009jf,Evans:2011pz}).%
\footnote{Alternatively, we may introduce
an SU$(2n)$ supersymmetric pure Yang-Mills sector. 
The SU$(2n)$ Yang-Mills theory possesses $\mathbb{Z}_{4nR}$ symmetry, which is spontaneously broken down to $\mathbb{Z}_{2R}$ by gaugino condensation~\cite{Witten:1982df,Veneziano:1982ah,Taylor:1982bp,Shifman:1987ia}.
}
In the following, we denote the
VEV of the superpotential in dynamical supersymmetry-breaking sector as
\begin{align}
W_\mathrm{dyn} \sim \Lambda_\mathrm{dyn}^3\ ,
\end{align}
where $\Lambda_{\mathrm{dyn}}$ 
denotes the dynamical scale of the hidden sector.

We assume that dynamical supersymmetry breaking occurs before inflation. 
Then, $X$ and $Y$ acquire a linear potential via supergravity effects,
\begin{align}
\mathit{\Delta} V_{X,Y} \sim \frac{1}{M_\mathrm{Pl}^2} W_\mathrm{dyn}^\dagger (v_X^2 X + v_Y^2 Y) + \hc
\end{align}
With these linear terms, $X$ and $Y$ are slightly displaced from their origins during inflation,
\begin{align}
\mathit{\Delta} X \sim -\frac{W_\mathrm{dyn}}{\kappa_X v_X^2}\ , \quad
\mathit{\Delta} Y \sim  \frac{W_\mathrm{dyn}}{(1-\kappa_{XY}) v_X^2}\ ,
\end{align}
where we have used the effective mass term in \cref{eq:VXY}. Consequently, the problematic $\mathbb{Z}_{4R}$ symmetry is broken at the onset of inflation if the displacements are sizable, i.e.,
\begin{align}
\label{eq:XYshifts}
\Delta X \gtrsim H_\mathrm{inf}\ , \quad \Delta Y \gtrsim H_\mathrm{inf}\ .
\end{align}
This is achieved for 
\begin{align}
\label{eq:req Lambda}
\Lambda_\mathrm{dyn}^3 \gtrsim \frac{1}{M_\mathrm{Pl}} v_X^4\ .
\end{align}
Once $X$ and $Y$ are displaced from their origins, no domain walls are formed after inflation, since $\mathbb{Z}_{4nR}$ symmetry is already broken at the onset.
Note also that $W_\mathrm{dyn}$ contribution does not displace $H$ and $\bar{H}$ from their origins, as they are protected by the $\uoneG$ symmetry.

For the reference value in \cref{eq:vXref}, this condition amounts to 
\begin{align}
\Lambda_\mathrm{dyn} \gtrsim 10^{13}\,\mathrm{GeV}\ .
\end{align}
Accordingly, the contribution of $W_\mathrm{dyn}$ to the gravitino mass is
\begin{align}
m_{3/2,\,\mathrm{dyn}}=
\frac{1}{M_\mathrm{Pl}^2}W_\mathrm{dyn}\gtrsim 10^{2}\,\mathrm{GeV}\ ,
\end{align}
which is subdominant compared to the gravitino mass from the inflaton sector in \cref{eq:m32 for n=3}. 
Since the required dynamical scale $\Lambda_\mathrm{dyn}$ is much larger than the Hubble scale during inflation, $\order{10^{10}}\,\mathrm{GeV}$, this is consistent with our assumption that $\mathbb{Z}_{4nR}$ symmetry breaks before inflation. 
Let us also note that \cref{eq:XYshifts} shifts the inflaton energy density only by $\order{H_\mathrm{inf}^4}$.
Thus, the domain wall problem
can be evaded with a reasonable parameter choice without modifying the inflaton dynamics.

\section{Reheating}
\label{sec:reheating}
\begin{figure}[t]
    \centering
\includegraphics[width=0.8\linewidth]{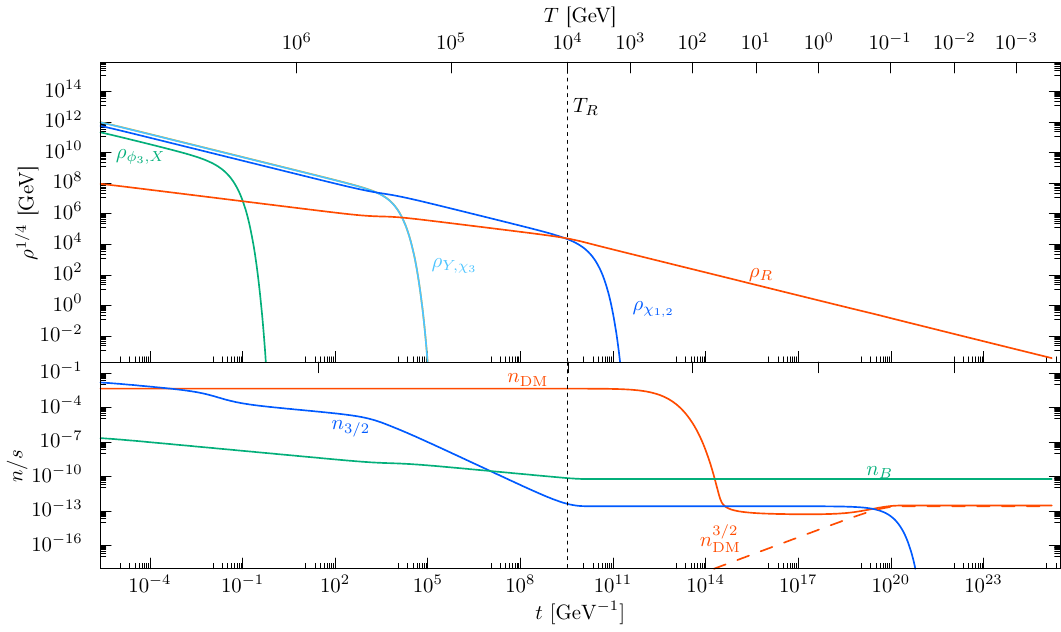}
\caption{Schematic time evolution of the energy densities (top) 
and yields (bottom)
in the early matter domination era after inflation as a function of the cosmic time.
We set $T_R = 10^4$\,GeV and $m_{3/2}=1$\,PeV, $m_\mathrm{LSP}\simeq 1$\,TeV. 
This figure depicts the period after the onset of oscillations for  $Y$  and  $\chi_{1,2,3}$.
The solid lines show the energy densities 
of $\phi_3+X$, $\chi_3+Y$, $\chi_1+\chi_2$, and the radiation $R$.
The reheating process 
is completed by the decay of $\chi_{1,2}$.
The other inflaton sector fields $\phi_3$, $X$, $\chi_3$ and $Y$ have decayed well before the completion of the reheating. 
The gravitinos are produced by the decays of $X$, $\phi_3$, $Y$ and $\chi_3$.
The baryon asymmetry is generated by the 
decay of $\chi_{1,2}$ into the right-handed neutrinos, which subsequently causes the non-thermal leptogeneisis. 
The LSP dark matter freezes out from the thermal bath at the temperature around 100\,GeV, which receives additional contribution from the decay of the gravitino at $T\sim 0.1$\,GeV. 
     }
    \label{fig:rhohistory}
\end{figure}
In this section, we discuss the reheating process of the inflaton sector.
The reheating process in the present model is somewhat complex, as seen below. For illustrative purposes, we show a schematic time  evolution of the  energy densities in \cref{fig:rhohistory}.

The decay rates of the inflaton sector fields will be explained in \cref{sec:decay}.
The typical time evolution is as follows.
After inflation, the inflaton $\phi_3$ begins oscillating while other fields are stabilized around their origins.
When the inflaton oscillation energy density decreases down to $v_Y^4$, other fields start to oscillate.
The reheating process takes place at much lower energy density determined by the decay rate of
the lightest inflaton sector fields, $\chi_{1,2}$.
The other inflaton sector fields $\phi_3$, $X$, $\chi_3$ and $Y$ have decayed into a pair of $\chi_{1,2}$'s well before the completion of the reheating. 

As an important note, 
the inflaton sector fields obtain large VEVs, and hence, 
the inflaton sector fields other than $\chi_{1,2}$ have sizable branching fraction into a pair of the gravitinos~\cite{Endo:2006zj,Nakamura:2006uc,Endo:2006tf,Kawasaki:2006gs,Dine:2006ii,Asaka:2006bv,Kawasaki:2006hm,Endo:2006qk,Nakayama:2012hy}.
As we will see in Sec\,\ref{sec:gravitino}, the abundance of non-thermally produced gravitinos puts a severe upper limit on the reheating temperature.

\subsection{High-Scale Supersymmetry}
\label{sec:MSSM}
Before discussing the thermal history in detail, we first review the setup of the supersymmetric standard model (SSM). 
Here, we assume the gravitino mass is of order PeV. 
Under this assumption, we outline the SSM within the high-scale supersymmetry framework~\cite{Hall:2011jd, Hall:2012zp, Nomura:2014asa, Ibe:2011aa, Ibe:2012hu, Arvanitaki:2012ps, ArkaniHamed:2012gw} characterized by a PeV-scale gravitino mass.

A primary advantage of a PeV-scale gravitino scenario is that it generates TeV-scale gaugino masses via anomaly-mediated supersymmetry breaking (AMSB), independent of specific details of dynamical supersymmetry-breaking sector~\cite{Randall:1998uk, Giudice:1998xp}.
This feature, which eliminates the need for a singlet supersymmetry-breaking field, is crucial in avoiding the cosmological Polonyi problem~\cite{Coughlan:1983ci,Goncharov:1984qm,Ellis:1986zt}. 
This should be contrasted with the gravity mediation models, where  the Polonyi problem persists even if supersymmetry breaking is dynamical and occurs through strong coupling~\cite{Ibe:2006am}.
In addition, it is also advantageous that the PeV-scale gravitino decays well before the Big Bang Nucleosynthesis (BBN), preserving successful BBN predictions.

Typical sfermion masses for a PeV-scale gravitino are of the same order as the gravitino mass. 
These heavy sfermions can easily explain the observed Higgs boson mass of $125$\,GeV~\cite{Okada:1990vk,Okada:1991jc,Ellis:1990nz,Haber:1990aw,Ellis:1991zd}. 
With these considerations, we adopt the following high-scale supersymmetry spectrum:
\begin{align}
\label{eq:PeVSUSY}
m_\mathrm{gaugino} = \order{1}\,\mathrm{TeV}, \quad m_\mathrm{sfermion} \simeq m_{3/2} = \order{100}\,\mathrm{TeV} - \order{1}\,\mathrm{PeV}\ .
\end{align}

A notable feature of the AMSB model is that the Wino, the SU(2)$_L$ gaugino in the SSM, typically becomes the lightest supersymmetric particle (LSP). 
In what follows, we consider Wino LSP as an example; however, the current inflation model is also compatible with a Higgsino LSP or a more general neutralino LSP with mixed Higgsino/Bino/Wino content.

\subsection{Decay of Inflaton Sector Fields and Reheating Process}
\label{sec:decay}
In the benchmark scenarios in \cref{sec:reference},
the masses of the fields in the inflaton sector are ordered as,
\begin{align}
    M_{X\phi_3} > M_{Y\chi_3} > 2\times M_{\chi_{1,2}} \ .
\end{align}
With this mass spectrum, 
the fields $X$, $\phi_3$, $Y$, and $\chi_3$ decay into $\chi_{1,2}$ via supersymmetric effective Yukawa interactions,
\begin{align}
    W_\mathrm{eff}
    =& \frac{3\zeta}{2}
    \frac{v_\mathrm{U(1)}^2}{M_\mathrm{Pl}^2}
    \phi_3 \chi_3^2
    - \frac{ 2\zeta v_\mathrm{U(1)}^2}{r_V^2(1+2 r_V^2)M_\mathrm{Pl}^2} \phi_3 \qty(\chi_1^2+
    \chi_2^2 )  - \frac{3\lambda_X r_V^4 v_\mathrm{U(1)}^4}{(1+2 r_V^2)M_\mathrm{Pl}^4}  X \qty(\chi_1^2+
    \chi_2^2 ) \cr
&   -  \frac{\zeta r_V(3+4 r_V^2)v_\mathrm{U(1)}^2}{\sqrt{2}(1+2 r_V^2)M_\mathrm{Pl}^2} \chi_3 \qty(\chi_1^2+
    \chi_2^2 )
- \frac{3\lambda_Y r_V^2 v_\mathrm{U(1)}^4}{(1+2 r_V^2)M_\mathrm{Pl}^4}  Y \qty(\chi_1^2+
    \chi_2^2 )\ ,
\end{align}
which originate from \cref{eq:Wpot}.
Here, we have shifted the chiral fields by the VEVs in 
\cref{eq:VEVs}, which eliminates tadpole terms in the superpotential.
Since $\zeta<\zeta_\mathrm{max}$ (see \cref{eq:zetamax}), we only consider terms up to the leading order of $\zeta$ in this section.

Additionally, the fields $X$, $\phi_3$, $Y$, and $\chi_3$ also decay into gravitino pairs, potentially leading to the gravitino problem~\cite{Endo:2006zj,Nakamura:2006uc,Endo:2006tf,Kawasaki:2006gs,Dine:2006ii,Asaka:2006bv,Kawasaki:2006hm,Endo:2006qk,Nakayama:2012hy}.%
\footnote{The decay of a scalar field into a superpartner and a gravitino pair can also contribute to the gravitino problem~\cite{Nilles:2001my,Ellis:2015jpg}. However, in this model, such decays are kinematically forbidden due to the small mass splitting within supermultiplets.}
However, because of the $\uone_H$ symmetry, $\chi_{1,2}$ do not decay into gravitino pairs but instead decay into the SSM sector, given the SSM fields possess the $\uone_H$ charges (see \cref{sec:B-L}). 
Consequently, the gravitino problem can be avoided with low reheating temperature. 
In the following, we investigate the reheating process. 

\subsubsection{Decay of \texorpdfstring{$\boldsymbol{X}$}{X} 
and 
\texorpdfstring{$\boldsymbol{\phi_3}$}{φ₃}
}
Let us first examine the decays of $X$ and $\phi_3$, which form a Dirac mass partner at $\zeta\to 0$. 
In the small $\zeta$ limit, 
the dominant decays proceed via the effective interaction terms, 
\begin{align}
W_\mathrm{eff} &=
M_{X\phi_3} X \phi_3  - \frac{1}{2} y_{X} X \qty(\chi_1^2 + \chi_2^2)\ ,\ \\
y_X &= \frac{6 \lambda_Y r_V^4 v_\mathrm{U(1)}^4}{(1 + 2 r_V^2) M_\mathrm{Pl}^4}\ .
\end{align}
The corresponding decay rates of $X$ and $\phi_3$ are given by%
\footnote{They also 
decay into $\chi_3$ with similar decay rate which does not change the following discussion.}
\begin{align}
\Gamma\qty(X \to \tilde{\chi}_{1,2} \tilde{\chi}_{1,2}) = \Gamma\qty(\phi_3 \to \chi_{1,2} \chi_{1,2}) \simeq \frac{y_X^2}{16\pi} M_{X\phi_3}\ ,
\end{align}
where the tilde denotes the fermionic component of a chiral superfield. In the benchmark scenarios discussed in \cref{sec:reference}, these decay rates are estimated as
\begin{align}\label{eq:GamX}
\Gamma\qty(X \to \tilde{\chi}_{1,2} \tilde{\chi}_{1,2}) = \Gamma\qty(\phi_3 \to \chi_{1,2} \chi_{1,2}) &\simeq 2.0 \times 10^{2}\,\mathrm{GeV}
\times \frac{\lambda_{X100}^{1/9}}{(1 + 0.76 r_\kappa^2 f_{T0.15})^2} \ .
\end{align}

As the scalar components of $X$ and $\phi_3$ acquire large VEVs, they can also decay into gravitino pairs. The corresponding decay rates are given by~\cite{Nakayama:2012hy},
\begin{align}
\label{eq:Gam32}
\Gamma(X \to \tilde{G} \tilde{G}) &\simeq
\frac{1}{64\pi} \qty(\frac{\langle \phi_3 \rangle}{M_\mathrm{Pl}})^2 \frac{M_{X\phi_3}^3}{M_\mathrm{Pl}^2}\ , \quad
\Gamma(\phi_3 \to \tilde{G} \tilde{G}) \simeq
\frac{1}{64\pi} \qty(\frac{\langle X \rangle}{M_\mathrm{Pl}})^2 \frac{M_{X\phi_3}^3}{M_\mathrm{Pl}^2}\ ,
\end{align}
where $\tilde{G}$ represents the gravitino. Here, it is assumed that the supersymmetry breaking field in the dynamical supersymmetry-breaking sector is heavier than the inflaton sector fields.
In the benchmark scenarios, the branching ratios into gravitinos are given by,
\begin{align}
\label{eq:BrX}
    \mathrm{Br}_{X,\tilde{G}}
    =\frac{\Gamma(X\to\tilde{G}\tilde{G})}{\Gamma(X\to \tilde{\chi}_{1,2}\tilde{\chi}_{1,2})}
    &\simeq
    2.6 \times 10^{-8}
  \times  (1+0.76 r_\kappa^2f_{T0.15})^2
\times \lambda_{X100}^{-8/9} 
       \ , \\
\label{eq:Brphi3}
\mathrm{Br}_{\phi_3,\,\tilde{G}}
    =\frac{\Gamma(\phi_3\to\tilde{G}\tilde{G})}{\Gamma(\phi_3\to \chi_{1,2}\chi_{1,2})}
&\simeq 1.5 \times 10^{-11}
    \times \lambda_Y^{2}
    \times \hat{\zeta}^{2}
    \times(1+0.76 f_{T0.15})^2  
   r_\kappa^{-14} f_{T0.15}^{-7}\times \lambda_{X100}^{-26/9}   \ . 
\end{align}

\subsubsection{Decay of \texorpdfstring{$\boldsymbol{Y}$}{Y}
and 
\texorpdfstring{$\boldsymbol{\chi_3}$}{χ₃}
}
Next, let us examine the decay rates of $Y$ and $\chi_3$, which form the Dirac mass partner 
in the small $\zeta$ limit.
As in the case of $X$ and $\phi_3$, 
decays of $Y$ and $\chi_3$ proceed through
\begin{align}
    W_\mathrm{eff}&= 
    M_{Y\chi_3} Y \chi_3  - \frac{1}{2}y_{Y} Y \qty(\chi_1^2+
    \chi_2^2 )\ ,\\
    y_Y&= \frac{6\lambda_Y r_V^2 v_\mathrm{U(1)}^4}{(1+2 r_V^2)M_\mathrm{Pl}^4}\ ,
\end{align}
resulting in
\begin{align}
\Gamma\qty(Y \to \tilde{\chi}_{1,2} \tilde{\chi}_{1,2}) = \Gamma\qty(\chi_3 \to \chi_{1,2} \chi_{1,2})  \simeq  
      \frac{y_Y^2}{16\pi} M_{Y\chi_3}\ .
\end{align}
Again, we have omitted the kinematical factor.
For the benchmark scenarios, the decay rates are given by,
\begin{align}
\label{eq:GamY}
 \Gamma\qty(Y \to \tilde{\chi}_{1,2} \tilde{\chi}_{1,2}) = \Gamma\qty(\chi_3 \to \chi_{1,2} \chi_{1,2})\simeq 1.1\times 10^{-2}\,\mathrm{GeV} 
    \times \lambda_Y^3 \times\frac{r_\kappa^{-9}f_{T0.15}^{-9/2}\lambda_{X100}^{-26/9}}{(1+0.76r_\kappa^2 f_{T0.15})^2} \ . 
\end{align}

The scalar components of $Y$ and $\chi_3$ also obtain large VEVs, and hence, 
their decay rates into gravitino pairs are given by
\begin{align}
\label{eq:Gam32II}
    \Gamma(Y\to \tilde{G}\tilde{G}) \simeq 
    \frac{1}{64\pi}\qty(\frac{\sqrt{2}\expval{H_1}}{M_\mathrm{Pl}})^2\frac{M_{Y\chi_3}^3}{M_\mathrm{Pl}^2}\ , \quad
   \Gamma(\chi_3\to \tilde{G}\tilde{G}) \simeq 
    \frac{1}{64\pi}\qty(\frac{\expval{Y}}{M_\mathrm{Pl}})^2\frac{M_{Y\chi_3}^3}{M_\mathrm{Pl}^2}\ .
\end{align}
For the benchmark scenarios, the branching ratios are given by 
\begin{align}
\label{eq:BrY}
   \mathrm{Br}_{Y,\,\tilde{G}}=\frac{\Gamma(Y\to\tilde{G}\tilde{G})}{\Gamma(Y\to \tilde{\chi}_{1,2}\tilde{\chi}_{1,2})}&\simeq
   1.2 \times 10^{-6}  \times
    r_\kappa^{-8}f_{T0.15}^{-4}\times(1+0.76 r_\kappa^2f_{T0.15})^2 \times 
    \lambda_{X100}^{-8/9} \ ,  \\
\label{eq:Brchi3}   
   \mathrm{Br}_{\chi_3,\,\tilde{G}}=\frac{\Gamma(\chi_3\to\tilde{G}\tilde{G})}{\Gamma(\chi_3\to \chi_{1,2}\chi_{1,2})}&\simeq  6.7 \times 10^{-8}
   \times   \hat{\zeta}^{2}\times
r_\kappa^{-8}f_{T0.15}^{-4}\times(1+0.76 r_\kappa^2f_{T0.15})^2  
\times    \lambda_{X100}^{-8/9}\ .
\end{align}

\subsubsection{Decay of \texorpdfstring{$\boldsymbol{\chi_1}$ and $\boldsymbol{\chi_2}$}{χ₁ and χ₂}}
\label{sec:B-L}
The decays of the lightest fields in the inflaton sector, $\chi_{1,2}$, require careful consideration. 
The present model possesses the $\uone_H$ symmetry under which $H$ and $\bar{H}$ transform as $e^{i\alpha}H$ and $e^{-i\alpha}\bar{H}$, respectively. 
At the supersymmetric vacuum in \cref{eq:VEVs}, this symmetry remains unbroken in combination with the $\uoneG$ gauge symmetry, manifesting as an SO(2) rotation of $(\chi_1, \chi_2)$ in \cref{eq:Wmassive}.Therefore, to induce the decay of $\chi_{1,2}$, it is necessary to break the $\uone_H$ symmetry or couple $\uone_H$-charged operators in SSM sector.

As one of such possibilities, we consider the following interaction term in the superpotential,
\begin{align}
\label{eq:Wdec}
    W_\mathrm{dec} = 
    \xi_{N\alpha \beta}\frac{(\epsilon H \phi H)}{M_\mathrm{Pl}^2}\bar{N}_{R\alpha}\bar{N}_{R\beta}\ .
\end{align}
Here, $\bar{N}_{R\alpha}$ ($\alpha=1,2,3$) are the right-handed neutrinos and $\epsilon$ is the two-dimensional anti-symmetric tensor, and $\xi_N$ is a coupling coefficient.
In this case, the $\uone_H$ charges of the SSM fields are assumed to be given by the $B-L$ charges.
This term is also consistent with the charge assignment in Table\,\ref{tab:symmetry}.
Note that this operator does not contribute to the right-handed neutrino masses since 
the $\uone_{B-L}$ symmetry,
which is given by a mixture of 
$\uone_H$ and $\uone_G$,
remains unbroken in the inflaton sector.

At around the supersymmetric vacuum
in \cref{eq:VEVs},
$\chi_{1,2}$ obtain decay operators in the superpotential,
\begin{align}    
\label{eq:Wdec2}
W_{\mathrm{dec}} &= -
\frac{1}{2}y_{\chi\alpha\beta}
    (\chi_1 + i \chi_2) \bar{N}_{R\alpha} \bar{N}_{R\beta}\ ,\\
    y_{\chi\alpha\beta}  &= \xi_{N\alpha\beta} \frac{(1+2r_V^2)^{1/2}v_\mathrm{U(1)}^2}{M_\mathrm{Pl}^2}\ .
\end{align}
Accordingly, the decay rates of $\chi_{1,2}$ are given by,
\begin{align}
    \Gamma(\chi_{1,2}\to \bar{N}_{R\alpha}\bar{N}_{R\beta}) = \frac{y_{\chi\alpha\beta}^2}{16\pi}M_{\chi_{1,2}} \ ,\quad (\alpha \ge \beta) \ .
\end{align}
For the benchmark points,  
\begin{align}
    \Gamma(\chi_{1,2}\to \bar{N}_{R\alpha}+\bar{N}_{R\beta}) \simeq 9.0 \times10^2\,\mathrm{GeV} \times
    \xi_{N\alpha\beta}^2\times \lambda_Y\times \hat{\zeta}\times r_\kappa^{-11}f_{T0.15}^{-11/2}\times (1+0.76r_\kappa^2 f_{T0.15})^2
    \times
    \lambda_{X100}^{-10/9}\ . 
\end{align}
Here, we have omitted the 
right-handed neutrino mass.
For the sake of reheating, we suppose $\xi_N$ are small so that 
the decay rates of $\chi_{1,2}$ are much smaller than those of $X$,
$\phi_3$, $Y$ and $\chi_3$.

Several comments are in order.
We assume that $\uone_{H}$ (and hence $\uone_{B-L}$) is a gauge symmetry, which undergoes spontaneous breaking in another sector which we do not discuss 
in this paper.
The spontaneous breaking of $\uone_{B-L}$ takes place 
before completion of the 
reheating.
This assumption is reasonable
since we require $M_R < M_{\chi_{1,2}}$, and hence, the $\uone_{B-L}$ breaking scale, $v_{B-L}$, is of $\order{10^8}$--$\order{10^{10}}$\,GeV (see \cref{eq:Mchi12}).
Note that the mixing
between $\uone_H$ and $\uone_G$
does not affect the mass spectrum or decay properties in the inflaton sector, provided the gauge coupling constant of $\uone_{H}$ is small.

\subsection{Post-Inflation Process}

\begin{figure}[t]
    \centering
        \centering
        \includegraphics[width=0.65\linewidth]{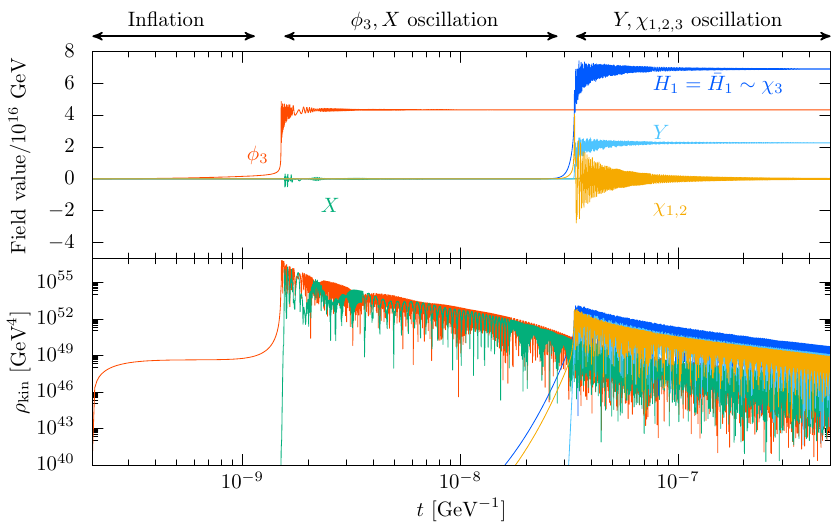}
    \caption{
    Example of the time evolution
    the inflaton sector fields (top) and the 
    kinetic energy (bottom) for BP$_1$.
 }
    \label{fig:rhooscillation}
\end{figure}
Now, let us examine the post-inflation process.
Following the end of inflation, $\phi_3$ begins oscillating around its VEV in \cref{eq:VEVs} ($\varphi_*$ in \cref{fig:potential}). 
The other fields remain stabilized at their origins due to the Hubble-induced masses, as discussed in \cref{sec:Stabilization}.
As the energy density of the $\phi_3$ oscillation, $\rho_{\phi_3}$, decreases to
\begin{align}
    \rho_{\phi_3} \lesssim v_Y^4\ ,
\end{align}
the fields $H$ and $\bar{H}$ become destabilized due to the negative Hubble mass terms and subsequently roll down to their potential minima.
At the same time, $X$ and $Y$ also start 
oscillation.
This post-inflation process induces the $\uone_G$ spontaneous breaking.
In \cref{fig:rhooscillation}, we illustrate a representative time evolution of the energy density during and after inflation.

While the dynamics is intricate, 
the inflaton sector ends up oscillating around the vacuum given by \cref{eq:VEVs}.
Through numerical analysis, 
we confirmed that all inflaton sector fields oscillate 
around their VEVs.
\Cref{fig:rhooscillation} shows an example of the time evolution 
of the inflaton sector fields. 
Note that the fraction of the oscillation energy of each inflaton sector field 
depends on their initial condition
after inflation.
After the onset of the $Y$, $\chi_{1,2,3}$ oscillation, the energy densities of $X$ and $\phi_3$ tend to be smaller than those of $Y$ and $\chi_{1,2,3}$.
On the other hand, the energy fraction of $\chi_{1,2}$ can be similar to $Y$  
as seen in \cref{fig:rhooscillation}.%
\footnote{The energy fractions are influenced by the Hubble fluctuations of the inflaton sector fields, resulting in variations from one Hubble patch to another after inflation. Consequently, these fractions are determined by ensemble averages.}
In the following, without detailing further, 
we treat the energy fractions 
as free parameters.

Since the decay rates of the inflaton sector fields are low, their decay processes occur long after the oscillatory phase.
The relatively short-lived fields, $X$, $\phi_3$, $Y$ and $\chi_3$, first decay into $\chi_{1,2}$. The long-lived $\chi_{1,2}$ decay into right-handed neutrinos via the decay operator \eqref{eq:Wdec2}.
The right-handed neutrinos promptly decay into SM particles through Yukawa interactions with the Higgs and lepton doublets in the SM.

It is noteworthy that the $\SU(2)$ gauge symmetry is broken down to the $\uone_G$ gauge symmetry at the onset of inflation, where $\phi_3 \neq 0$.
Therefore, no magnetic monopoles associated with the $\SU(2)$ breaking survive in the observable Universe.
In contrast, the $\uone_G$ symmetry breaking occurs after inflation, once the energy density of the Universe drops below $v_Y^4$. 
The $\uone_G$ symmetry breaking leads to formation of the cosmic string network with the correlation length of the horizon scale at that time.
The evolution of the cosmic strings will be further discussed in \cref{sec:CosmicString}.

\subsection{Non-Thermal Gravitino Production}
\label{sec:gravitino}
As we have discussed, 
$X$, $\phi_3$, $Y$
and $\chi_3$ decay into pairs of gravitinos,
which takes place in the oscillating phase of the inflaton sector.
Considering that $\chi_{1,2}$ are the last to decay in the inflaton sector, the gravitino yield is given by
\begin{align}
\label{eq:Y32}
Y_{3/2}:=  \frac{n_{3/2}}{s} \simeq \bar{\kappa}\times \frac{3}{4} \times \qty(    
\mathrm{Br}_{X,\tilde{G}} \frac{r_X T_R}{M_{X\phi_3}}+
\mathrm{Br}_{\phi_3,\tilde{G}} \frac{r_{\phi_3}T_R}{M_{X\phi_3}}
+\mathrm{Br}_{Y,\tilde{G}} \frac{r_Y T_R}{M_{Y\chi_3}}
+\mathrm{Br}_{\chi_3,\tilde{G}} \frac{r_{\chi_3} T_R}{M_{Y\chi_3}}
    )\ .
\end{align}
where $s$ denotes the entropy density 
of the SSM sector after reheating.
The parameters $r_A$ $(A=X,\phi_3,Y,\chi_3)$ are the 
energy fractions of the oscillation energy density
$\rho_A$ well before their decay times,
\begin{align}
    r_{A} = \frac{2\rho_A}{\rho_{\chi_1}+\rho_{\chi_2}}\ .
\end{align}
The reheating temperature 
$T_R$ is defined by the decay rate of $\chi_{1,2}$,
\begin{align}
\label{eq:TR}
    T_R = 
    \qty(\frac{10}{\pi^2 g_*(T_R)}\Gamma_{\chi_{1,2}}^2M_\mathrm{Pl}^2)^{1/4}\ .
\end{align}
The factor $\bar{\kappa} \simeq 1.6$ denotes the universal fudge factor to calibrate the instantaneous decay approximation at $T_R$ to a slowly decaying process \cite{Ibe:2005jf}.

Note that the non-thermally produced gravitino abundance is suppressed more for lower reheating temperature. 
This should be contrasted with the non-thermal gravitino abundance in Ref.\,\cite{Nakayama:2012hy},
where the gravitino abundance is suppressed for a higher reheating temperature.
This essential difference is that the reheating is completed by
the decay of $\chi_{1,2}$, 
which do not decay into the gravitinos.

\begin{figure}[t]
    \centering
    \begin{subfigure}{0.45\textwidth}
            
        \centering
        \includegraphics[width=\linewidth]{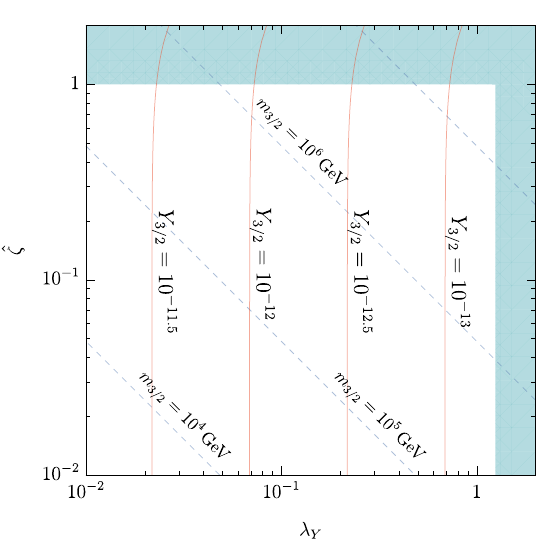}
        \caption{BP$_1$ with $T_R =10^{3.5}$\,GeV}
    \label{fig:subfig1}
    \end{subfigure}
    \hfill
    \begin{subfigure}
    {0.45\textwidth}
  \centering\includegraphics[width=\linewidth]{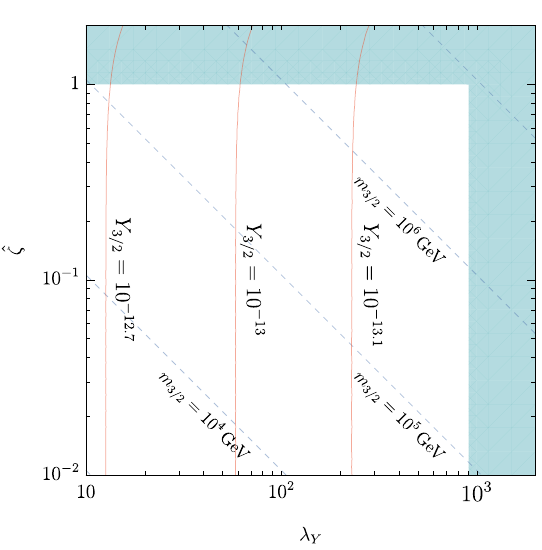}
         \caption{BP$_2$ with $T_R =10^5$\,GeV}           \label{fig:subfig2}
    \end{subfigure}
    \caption{The gravitino abundance produced by the inflaton sector fields decay in \cref{eq:Y32} 
    for the benchmark scenarios in \cref{eq:benchmark}
    (red lines).
    Here, we take $r_A = 1$ $(A=X,\phi_3,Y,\chi_3)$.
   The gravitino abundance is proportional to $T_R$.
   The blue (dashed) contours show the gravitino masses.
    The horizontal shaded regions are excluded by the 
condition in \cref{eq:smooth}.
The vertical shaded regions are excluded by the condition 
$v_Y^4/v_X^4<0.05$.
    }
    \label{fig:Y32}
\end{figure}

\Cref{fig:Y32} shows 
the yield of the gravitino produced by the decay of the inflaton sector fields for the benchmark scenarios in \cref{eq:benchmark} with 
$r_A = 1$ $(A=X,\phi_3,Y,\chi_3)$.
The horizontal shaded regions are excluded by the condition for smooth $\uone_G$ symmetry breaking in \cref{eq:smooth}.
The vertical shaded regions are excluded by the condition 
$v_Y^4/v_X^4<0.05$.
In the parameter range in \cref{fig:Y32}, 
the conditions in 
\cref{eq:XandYVEVs}
and the spectrum ordering 
in \cref{eq:spectrum}
are satisfied.
For $r_A = 1$ $(A=X,\phi_3,Y,\chi_3)$, the non-thermal gravitino production is dominated by the decay of $Y$ (see \cref{eq:BrY}).
Hence, the gravitino abundance is roughly proportional to $r_Y$
in the parameter range we 
are interested in.

Now, let us summarize constraints on the gravitino abundance.
For the SSM spectrum in \cref{eq:PeVSUSY}, the gravitino decays into 
a gauge boson-gaugino pair in the SSM.
The corresponding lifetime is given by 
\begin{align}
    \tau_{3/2}\simeq 3\times 10^{-5}\,\mathrm{sec}\times
    \qty(
    \frac{10^6\,\mathrm{GeV}}{m_{3/2}}
    )^3\ ,
\end{align}
or the decay temperature,
\begin{align}
    T_{3/2} \simeq 
    0.2\,\mathrm{GeV} \times g_*( T_{3/2})^{-1/4}\qty(\frac{m_{3/2}}{10^6\,\mathrm{GeV}})^{3/2}
\end{align}
(see e.g., Ref.\,\cite{Moroi:1995fs} for review).
Thus, 
 for $m_{3/2}\gtrsim 100$\,TeV, the gravitino decays well before the big-bang nucleosynthesis (BBN), and hence, the BBN does not put constraints on $Y_{3/2}$~\cite{Kawasaki:2008qe}.

On the other hand, assuming the LSP in the SSM sector is stable,
the decay of the gravitino contributes to the LSP dark matter abundance.
Accordingly, the gravitino abundance is constrained as%
\footnote{For the LSP mass in the TeV range, their freeze-out temperature is above $\order{10}$\,GeV.
Thus, the LSP's produced by the gravitino decay do not annihilate with each other for $m_{3/2}=\order{10^6}$\,GeV.}
\begin{align}
    m_\mathrm{LSP}\times \qty(Y_{3/2} + Y_\mathrm{th}) < 4.3 \times 10^{-10}\,\mathrm{GeV}
    \times \qty(\frac{\Omega_\mathrm{DM}h^2}{0.12})\ ,
\end{align}
where $Y_\mathrm{th}$ is the LSP abundance determined by the thermal freeze-out process.
Thus, for the LSP 
whose thermal relic abundance is below the observed value, the gravitino abundance is required to be in the range of
\begin{align}
\label{eq:Y32 upper}
Y_{3/2} \lesssim 4 \times 10^{-13}   \times \qty(\frac{\Omega_\mathrm{DM}h^2}{0.12})
\qty(\frac{1\,\mathrm{TeV}}{m_\mathrm{LSP}})\ .
\end{align}
\Cref{fig:Y32} shows that 
for the BP$_1$ region,
the constraint in \cref{eq:Y32 upper} can be satisfied for $r_YT_R = \order{10}\,$TeV or below, while it cannot be satisfied for $r_YT_R\gtrsim 10^{5}$\,GeV.
For the BP$_2$ region,
the constraint from the gravitino 
is much weaker.

\subsection{Non-Thermal Leptogenesis}
The decay operator in \cref{eq:Wdec} facilitates non-thermal (supersymmetric) leptogenesis~\cite{Fukugita:1986hr} through the decay of $\chi_{1,2}$. Upon production by $\chi_{1,2}$ decay, the right-handed neutrinos decay into the up-type Higgs doublet $H_u$ and the left-handed lepton doublets $\ell$. The CP violation in neutrino Yukawa interactions induces lepton number asymmetry \cite{Fukugita:1986hr,Covi:1996wh,Buchmuller:1997yu} as
\begin{align}
    \epsilon_\alpha \equiv \frac{\Gamma(\bar{N}_{R\alpha} \to H_u + \ell) - \Gamma(\bar{N}_{R\alpha} \to H_u^\dagger + \ell^\dagger)}{\Gamma(\bar{N}_{R\alpha} \to H_u + \ell) +
    \Gamma(\bar{N}_{R\alpha} \to H_u^\dagger + \ell^\dagger)}\ ,
\end{align}
where $\alpha=1,2,3$ represents the generation of right-handed neutrinos and flavor indices of lepton doublets are suppressed. For detailed expressions of $\epsilon_\alpha$ involving neutrino Yukawa couplings and right-handed neutrino masses $M_{R\alpha}$, refer to Ref.\,\cite{Covi:1996wh}.

The resultant baryon asymmetry from non-thermal leptogenesis is given by~\cite{Lazarides:1990huy,Ibe:2005jf},
\begin{align}
    \frac{n_B}{s} &\simeq \bar{\kappa}\times C_\mathrm{sph}\times 
    \frac{3}{4}\frac{T_R}{M_{\chi_{1,2}}} 
      \sum_{\alpha=1}^{3} 2\epsilon_{\alpha} \mathrm{Br}_{\alpha}\ .  
\end{align}
Here, $\mathrm{Br}_\alpha$ is the branching fraction of $\chi_{1,2}$ to $\bar{N}_{R\alpha}$, $C_\mathrm{sph}$ is the $L$
to $B$ conversion factor 
via the sphaleron process, and $\bar{\kappa}\simeq 1.6$ corrects the instantaneous decay approximation of $\chi_{1,2}$ at $T_R$. Since $T_R \sim 10^4$\,GeV, we may reasonably take $M_{R_\alpha} \gg T_R$ even for $M_{R_\alpha} < M_{\chi_{1,2}}$, and thus, the lepton asymmetry from heavier neutrinos is not washed out by lighter ones.

With a reheating temperature below the sfermion masses (\cref{sec:MSSM}), the radiation dominated era's thermal bath comprises SM particles and gauginos. Hence, the sphaleron conversion factor is $C_\mathrm{sph}=28/79$~\cite{Khlebnikov:1988sr,Harvey:1990qw}, yielding
\begin{align}
    \frac{n_B}{s} &\simeq 8.2\times 10^{-13}\times
    \frac{1}{\sin^2\beta}\qty(\frac{T_R}{10^4\,\mathrm{GeV}})
    \qty(\frac{m_{\nu_3}}{0.05\,\mathrm{eV}})
    \qty(\frac{2 M_{R1}}{M_{\chi_{1,2}}})
    \times \sum_{\alpha=1}^3 
    \mathrm{Br}_\alpha
    \frac{\epsilon_\alpha}{\epsilon_\mathrm{DI}}\ .   
\end{align}
where $\beta$ is the 
arctangent of the ratio of the two Higgs VEVs in the SSM.
The observed Higgs mass 
can be achieved
for $\tan\beta= \order{1}$ 
and the sfermion masses around $\order{10^6}$\,GeV. Assuming normal ordering of active neutrino masses, $m_{\nu 3}$ is the heaviest neutrino mass. We normalize lepton asymmetry by the Davidson-Ibarra bound on $\epsilon_1$~\cite{Davidson:2002qv},
\begin{align}
\epsilon_\mathrm{DI} := \frac{3}{8\pi}
\frac{M_{R1}}{\langle H_u^0\rangle^2} m_{\nu_3}\ .
\end{align}
Here, we order the right-handed neutrinos as $M_{R_1} < M_{R_2} < M_{R_3}$. 

As shown in Ref.\,\cite{Asaka:2002zu} (see also Ref.\,\cite{Zhang:2023oyo} for recent work), the CP asymmetry for heavier right-handed neutrinos may exceed $\epsilon_\mathrm{DI}$. For example, with $M_{2}/M_{1}\simeq M_{3}/M_{2} \simeq 2$, we can achieve $\epsilon_{2}/\epsilon_\mathrm{DI}=\order{10^{2\mathchar`-3}}$ or even more while preserving active neutrino masses and mixing through the seesaw mechanism~\cite{Minkowski:1977sc,Yanagida:1979as,*Yanagida:1979gs,Gell-Mann:1979vob,Glashow:1979nm,Mohapatra:1979ia}. Consequently, the observed baryon asymmetry $n_B/s \simeq 8.7 \times 10^{-11}$ can be explained by non-thermal leptogenesis for $T_R \sim 10^4$\,GeV.

\section{Cosmic String and Gravitational Wave Signal}
\label{sec:CSGWS}
As discussed in the previous section, 
the $\SU(2)$ symmetry is broken during inflation, while the $\uone_G$ symmetry breaking occurs after inflation. Consequently, the string network associated with the breaking of the $\uone_G$ symmetry is formed. In this section, we explore the characteristics of the cosmic strings in this model and the GW signals produced by this string network.

Before proceeding further, it is important to note theoretical uncertainties 
in the GW spectrum generated by the metastable 
string.
In models where the monopole at the endpoint of the metastable string 
has unconfined magnetic flux associated with an unbroken $\uone$ gauge symmetry, 
the string segments dissipate most of their energy by emitting massless gauge bosons shortly after their formation~\cite{Berezinsky:1997kd}.
Conversely, if the monopole flux is confined, the string segments lose energy solely via GW emission~\cite{Martin:1996cp}, giving additional contribution to the GW spectrum.

As discussed in \cref{sec:reheating}, 
we assume the model possesses the U$(1)_{H}$ gauge symmetry, leading to 
the $\uone_{B-L}$ symmetry via the 
VEVs of $H_1$ and $\bar{H}_1$.
Since $\uone_{B-L}$ arises as a mixture of $\uone_G$ and $\uone_H$, 
the monopoles attached to the string endpoints would have unconfined magnetic flux of $\uone_{B-L}$ if the metastable strings break in the $\uone_{B-L}$ symmetric phase.
However, in this model, the $\uone_{B-L}$ symmetry breaking occurs prior to string breaking, meaning that the monopoles at the string endpoints do not carry unconfined magnetic flux.

Nonetheless, the rate of energy dissipation of string segments due to interactions between the $\uone_{B-L}$ magnetic flux near the string endpoints and $\uone_{B-L}$ charged particles in the thermal bath remains an open question.
The energy loss due to the emission of $\uone_{B-L}$ gauge bosons from accelerated monopoles also requires further examination.
These issues are outside the scope of this paper. Instead, following Ref.~\cite{NANOGrav:2023hvm}, we consider two scenarios: one 
where only loops contribute to the GW spectrum (META-L),
and another where both loops and segments contribute to the GW spectrum (META-LS).

Let us also note that the mixing between $\uone_G$ and $\uone_H$ does not affect the formation of the $\uoneG$ string network. 
In the presence of the mixing, the winding number of the $\uoneG$ string  is given by $\pi_1\qty[\uoneG\times \uone_{H}/\uone_{B-L}] = \mathbb{Z}$.
The cosmic strings associated with the $\uone_{B-L}$ symmetry breaking in a separate sector, on the other hand, evolve into an independent network with a significantly lower string tension than that of the $\uoneG$ strings.

\subsection{Cosmic String Configuration}
\label{sec:CosmicString}
Let us discuss the local string formed at 
the $\uone_G$ gauge symmetry breaking after inflation.
Since $\SU(2)$ is spontaneously broken 
by $\langle \phi_3\rangle$ during inflation,
the effective theory relevant for the cosmic string configuration consists of the $\uone_G$ gauge boson $A^3_\mu$, four charged complex scalar fields $H_{1,2}$ and $\bar{H}_{1,2}$, and the gauge singlets $X$ and $Y$.

As discussed in \cref{sec:domain wall}, the $\mathbb{Z}_{4R}$ symmetry 
is broken by $W_\mathrm{dyn}$ well before inflation.
We assume, without loss of generality, that our Universe is in the domain 
where the vacuum is given by \cref{eq:VEVs},
i.e., $H_1=\bar{H}_1\neq 0$ while $H_2 = \bar{H}_2=0$.
Accordingly, the string configuration in our Universe is the one 
where the complex phases of $H_1$ and $\bar{H}_1$ wind, while $H_2$ and $\bar{H}_2$ remain zero.
Additionally, since 
the $D$-term potential 
is steeper 
compared with other contributions to the scalar potential, 
we focus on the $D$-flat direction,
\begin{align}
    |H_1| = |\bar{H}_1| \ .
\end{align}

\begin{figure}[tb]
\centering
\begin{minipage}{0.48\linewidth}
\includegraphics[width=\linewidth]{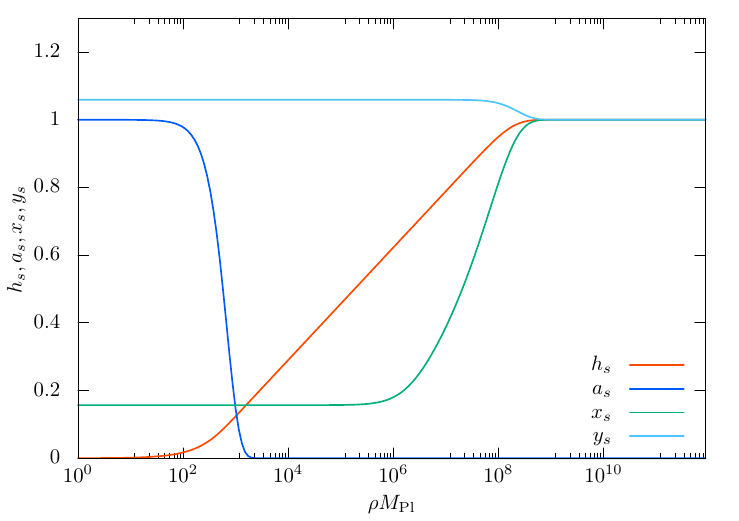}
\end{minipage}
\begin{minipage}{0.48\linewidth}
\includegraphics[width=\linewidth]{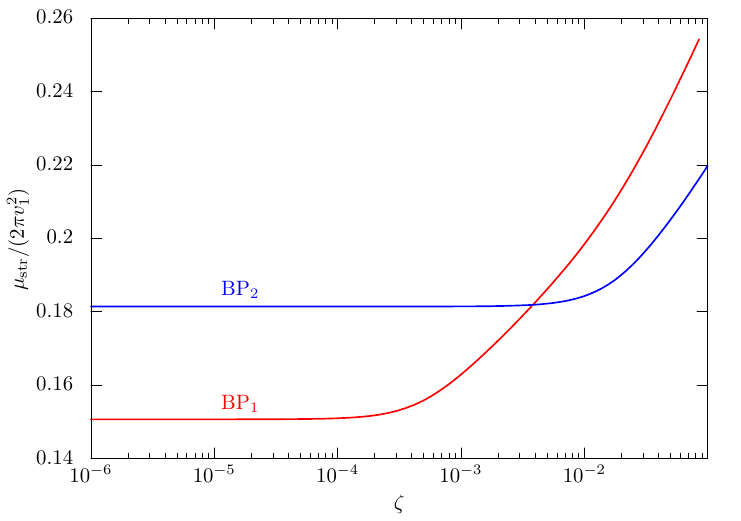}
\end{minipage}
\caption{Left) String profile functions 
as functions of the
radius of the cylindrical coordinate,
$\rho$,
in units of $M_\mathrm{Pl}^{-1}$.
Right) The string tension coefficient $f_T$ as a function of 
$\zeta$.
In the figure, we take
$\lambda_X = 100$ and $\lambda_Y = 0.1$.
}
\label{fig:string_configuration}
\end{figure}

The static string solution along the $x^3$-axis takes the form (see, e.g., Ref.\,\cite{Vilenkin:2000jqa})
\begin{align}
\label{eq:string ansatz1}
H_1&=\bar{H}_1^\dagger  =v_\mathrm{U(1)}\times h_s(\rho)e^{-i n_w\azimuth}\ , \\
\label{eq:string ansatzX}
X&=\langle X \rangle \times
x_s(\rho) \ ,  \\
\label{eq:string ansatzY}
Y &= \langle Y \rangle \times y_s (\rho)\ , \\
\label{eq:string ansatz2}
A^{a=3} &= -2n_w \times (1-a_s(\rho))  d\azimuth
\end{align}
where $n_w \in \mathbb{Z}$ is the winding number of the string, and $h_s(\rho)$, $x_s(\rho)$, $y_s(\rho)$ and $a_s(\rho)$ are the profile functions.
We use the cylindrical coordinates along the $x^3$-axis
with the azimuthal angle
$\azimuth$ and the radial coordinate $\rho$.
The boundary conditions for the profile functions are specified as follows:
\begin{align}
h_s(\rho)\rightarrow 0\ , ~(\rho\rightarrow0)&\ ,~~~~~h_s(\rho)\rightarrow 1\ , ~(\rho\rightarrow \infty)\ ,\\
a_s(\rho)\rightarrow 1\ , ~(\rho\rightarrow 0)&\ ,~~~~~a_s(\rho)\rightarrow 0\ , 
~(\rho\rightarrow \infty)\ , \\
x_s'(\rho)\rightarrow 0\ , ~(\rho\rightarrow 0)&\ ,~~~~~x_s(\rho)\rightarrow 1 \ ,
~(\rho\rightarrow \infty)\ , \\
y_s'(\rho)\rightarrow 0\ , ~(\rho\rightarrow 0)&\ ,~~~~~y_s(\rho)\rightarrow 1\ , 
~(\rho\rightarrow \infty)\ . 
\end{align}
\Cref{fig:string_configuration} shows 
the profile functions for the reference parameter set in \cref{sec:reference}.

In the right panel of \cref{fig:string_configuration},
we show the string tension as a function of $\zeta$ for $n_w=1$.
The figure indicates that the string tension depends on $\zeta$.
For the reference value $\zeta=\order{10^{-4}}$, 
we find $f_T \sim 0.15$.
For an analytic estimate of the string tension
in the small $\zeta$ limit, see \cref{sec:tension}.

\subsection{GW Spectrum from Metastable Cosmic Strings}
Now let us discuss the stochastic GW spectrum from the metastable strings in the present model.
As discussed in \cref{sec:inflation}, 
the $\SU(2)$ symmetry breaking is expected to occur at $\order{10^{16}}$\,GeV to reproduce the observed curvature perturbations.
The corresponding string tension is 
given by \cref{eq:U1reference},
yielding
\begin{align}
G_\mathrm{N}\mu_\mathrm{str} \simeq 10^{-5}\ ,
\end{align}
for the benchmark scenarios.

One important note is that the GWs emitted from the cosmic string network during the radiation dominated (RD) era exhibit a flat spectrum at high frequencies,
\begin{align}
\label{eq:plateau}
\Omega_\mathrm{GW} |_\mathrm{RD}  \sim 10^{-7}  \times
    \qty( \frac{G_{\mathrm{N}} \mu_{\mathrm{str}}}{10^{-5}} )^{1/2} \ .
\end{align}
The flat spectrum in \cref{eq:plateau} is derived based on the velocity-dependent one scale (VOS) model~\cite{Martins:1996jp,Martins:2000cs} (see also \cref{sec:GW}).%
\footnote{For the VOS model parameters, see \cref{eq:VOS} (see Ref.\,\cite{Gouttenoire:2019kij}).}
In \cref{eq:plateau}, only GW emission from cusps on the string loops is considered~\cite{Blanco-Pillado:2017oxo}.
Consequently, the flat spectrum for $G_\mathrm{N}\mu_\mathrm{str}\simeq 10^{-5}$ in \cref{eq:plateau} is in strong tension with the stochastic GW constraints set by the LIGO–Virgo–KAGRA (LVK) collaboration~\cite{KAGRA:2021kbb},
\begin{align}
\label{eq:LVK}
\Omega_\mathrm{GW} 
\leq 1.7 \times 10^{-8}  
\quad \mbox{at} \, \quad
f_\mathrm{LVK} = 25\,\mathrm{Hz} \ .
\end{align}

Interestingly, the LVK constraint can be alleviated 
for a low reheating temperature where the early matter domination (MD) lasts longer (see \cref{fig:rhohistory}).
To see this, consider the relation between the observed frequency and the cosmic temperature at the GW emission $T_\mathrm{emit}$,
\begin{align}
\label{eq:fk}
    f^{(k)}(T_\mathrm{emit}) \sim 5\,\mathrm{Hz} \times k 
      \qty( \frac{10^{-5}}{G_{\mathrm{N}} \mu_{\mathrm{str}}} )
      \qty( \frac{T_\mathrm{emit}}{10^4\,\mathrm{GeV}} ) \ ,
\end{align}
for the radiation dominated era.
Here, $k$ denotes the harmonic mode with the emitted frequency $f_\mathrm{emit} = 2k/\ell$ 
from a loop of length $\ell$ at cosmic time $t_\mathrm{emit}$.
\Cref{eq:fk} is derived by using the dominant GW emission time
\begin{align}
\label{eq:dominantemission}
    t_\mathrm{emit} \simeq \frac{\ell}{2 \Gamma G_{\mathrm{N}} \mu_\mathrm{str}} \ ,
\end{align}
where $\Gamma\simeq 50$ is the total GW emission efficiency from the string network.
Thus, for a low reheating temperature $T_R \lesssim 10^4$\,GeV, the GW frequency around the LVK constraint turns out to be emitted during the early MD era for $k=\order{1}$ and $G_{\mathrm{N}} \mu_\mathrm{str} = 10^{-5}$.

\begin{figure}
    \centering
    \includegraphics[width=0.45\linewidth]{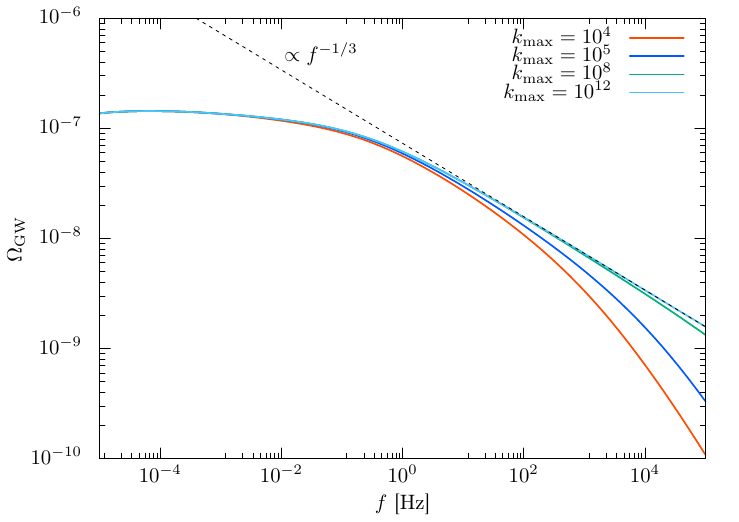}
    \caption{The high-frequency behavior of the GW spectrum for $G_{\mathrm{N}}\mu_\mathrm{str}=10^{-5}$, $\sqrt{\kappa}=7.7$, $T_R = 10^{4}$\,GeV, and various values of
    the maximum harmonic mode $k_\mathrm{max}$.}
    \label{fig:GW45}
\end{figure}

The GW spectrum of the frequencies emitted during the early MD era is suppressed by a factor of $f^{-1/3}$  on the high-frequency side (see \cref{eq:OmegaKMD}),
\begin{align}
    \Omega_\mathrm{GW}|_\mathrm{MD}  \propto f^{-1/3} \ . 
\end{align}
This suppression arises due to the faster cosmic expansion rate than in the RD era, which dilutes the number density of produced loops.
In addition, in the MD era, the VOS model predicts a longer correlation length between long strings than that in the RD era which reduces the string loop production rate. Consequently, the GW production efficiency is suppressed by a factor of $\order{0.1}$ compared with that in the RD era  (see \cref{eq:Ceff}).

\Cref{fig:GW45} shows the high-frequency behavior of the GW spectrum
for 
$G_\mathrm{N}\mu_\mathrm{str}=10^{-5}$, $\sqrt{\kappa}=7.7$, 
$T_R = 10^4\,$GeV,
and various maximum harmonic modes 
$k_\mathrm{max}$ of the summation (see \cref{eq:OmegaGW}).
Here, we take the parameters for the string network as
\begin{align}
    \label{eq:VOS}
    \tilde{c}=0.23\ , \quad \alpha = 0.1\ , \quad \mathcal{F}_\alpha = 0.1\ ,
    \quad \gamma_\alpha = \sqrt{2}\ ,
    \quad \Gamma=50 \ ,
\end{align}
(see also \cref{sec:GW}).
We only include the GW emission from cusps on the string loops.
The figure shows that the GW energy density spectrum
exhibits $f^{-1/3}$ slope for $f \gg f^{(1)}(T_R)$ with $f^{(1)}\simeq 5$\,Hz.

\begin{figure}[t]
    \centering
    \begin{subfigure}{0.45\textwidth}           
        \centering\includegraphics[width=\linewidth]{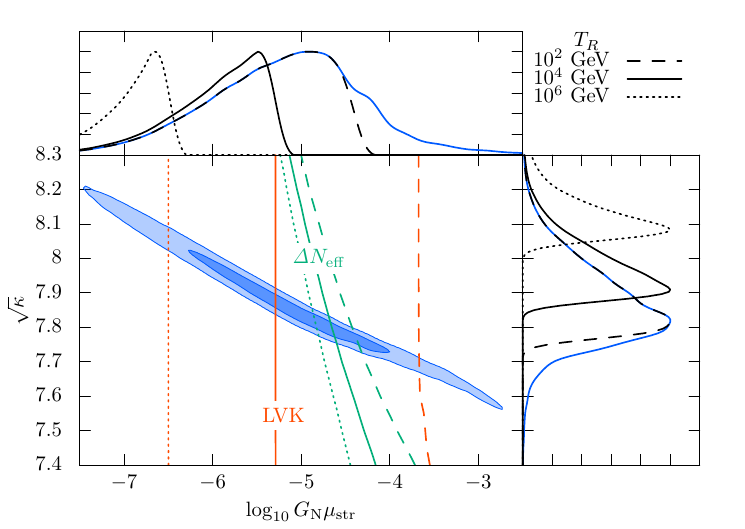}
        \caption{META-L ($\tilde{c}=0.23$)}
        \label{fig:META-L1}
    \end{subfigure}
    \begin{subfigure}{0.45\textwidth}
        \centering\includegraphics[width=\linewidth]{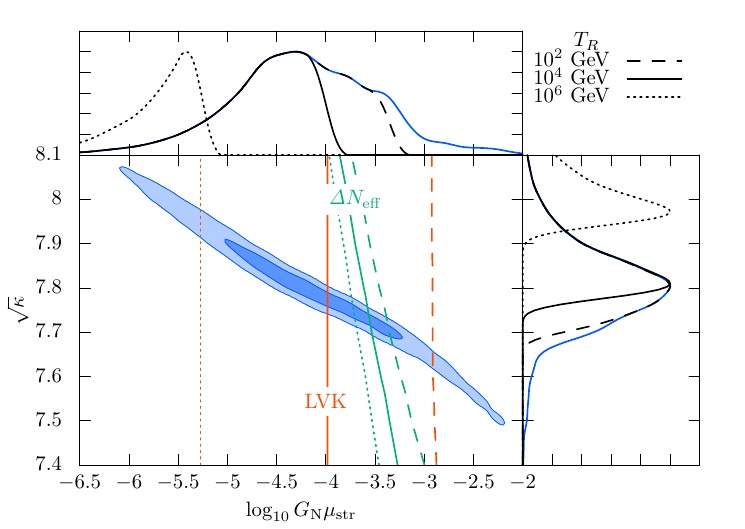}
        \caption{META-L ($\tilde{c}=0.57$)}           
        \label{fig:META-L2}
    \end{subfigure}
    \caption{The posterior distributions 
        of $\log_{10}G_\mathrm{N}\mu_\mathrm{str}$ 
and $\sqrt{\kappa}$ for the META-L model calculated with \texttt{PTArcade} and \texttt{Ceffyl}~\cite{andrea_mitridate_2023,Mitridate:2023oar,Lamb:2023jls}. 
    The panel (a) is for the VOS parameters \modifiedat{JCAP2}{calibrated} for the Nambu-Goto simulation. 
    The panel (\modifiedat{JCAP2}{b}) is for the VOS parameters calibrated for the Abelian-Higgs field theoretic simulation.
    The blue shaded regions indicate the 68\% (darker) and 95\% (lighter) Bayesian credible intervals. 
The red lines represent the LVK constraints on the GW spectrum; regions to the right are excluded.
The green lines show the CMB constraint on the effective GW energy density; regions to the right are excluded.
The black line in the one-dimensional posterior distribution panels represents the posterior distribution with LVK and CMB constraints applied.
The dotted, solid, and dashed lines correspond to $T_R = 10^6$\,GeV, $T_R = 10^4$\,GeV, and $T_R = 100$\,GeV, respectively.
   }
       \label{fig:META-L}
\end{figure}

In the rest of this section, we adapt
the Bayesian analysis for the preferred ranges of $G_\mathrm{N}\mu_\mathrm{str}$ and $\sqrt{\kappa}$
in Ref.\,\cite{NANOGrav:2023hvm} to our model.
First, we consider the case where 
the GWs are dominated by the cusp contributions on the string loops i.e., META-L. 
\Cref{fig:META-L} displays the 1D and 2D posterior distributions for $\log_{10}G_\mathrm{N}\mu_\mathrm{str}$ and $\sqrt{\kappa}$. 
\Cref{fig:META-L1}
corresponds to the loop-production coefficient $\tilde{c}\simeq 0.23$, which is calibrated to match Nambu-Goto simulations~\cite{Martins:2000cs}, which reproduces the result in Ref.\,\cite{NANOGrav:2023hvm}.
\Cref{fig:META-L2} corresponds to the loop-production coefficient $\tilde{c}\simeq 0.57$ which is calibrated to Abelian-Higgs simulations~\cite{Martins:2000cs}.
The latter case results in a lower GW signal for the same $G_\mathrm{N}\mu_\mathrm{str}$ due to longer correlation length between the long strings.
Note that the early MD era
only affects the GW spectrum in the high-frequency region, and hence, does not affect the GW spectrum in the PTA signal region.

In the 2D distributions, shaded regions denote 68\% (darker) and 95\% (lighter) Bayesian credible regions. The red lines indicate the LVK constraints on the GW spectrum in \cref{eq:LVK}; 
regions to the right of these lines are excluded. The solid line represents the constraints for $T_R = 10^4$\,GeV while the dashed and the long-dashed lines correspond to $T_R = 10^6$\,GeV and $T_R = 10^2$\,GeV, respectively.
This figure indicates that lower reheating temperatures weaken the LVK constraint, making both BP$_{1}$ and BP$_{2}$ scenarios feasible for $T_R \lesssim 10^3$\,GeV with $\tilde{c}\simeq 0.23$
and for $T_R \lesssim 10^4$\,GeV with $\tilde{c}\simeq 0.57$.
Also, the green lines represent the  constraints from CMB observations of the 
effective neutrino numbers $N_\mathrm{eff}$ (see also Ref.\,\cite{Servant:2023tua});  regions to the right of these lines are excluded.%
\footnote{
\modifiedat{JCAP2}{When cosmic strings survive until the time of the CMB or later (e.g., $\sqrt{\kappa} \gtrsim 8.5$), they affect the anisotropy of the CMB. These effects lead to a constraint of $G_\mathrm{N}\mu_{\text{str}} \gtrsim 10^{-7}$ from CMB observations~\cite{Planck:2013mgr}. Since our model prefers $\sqrt{\kappa} < 8$, this constraint is irrelevant.
}}
Here, we imposed $\mathit{\Delta}N_\mathrm{eff}\lesssim 0.3$ as the upper limit~\cite{Planck:2018vyg}. 
This constraint is also weakened for lower reheating temperatures due to suppressed high-frequency GWs.

\begin{figure}[t]
    \centering
    \begin{subfigure}{0.45\textwidth}    \centering\includegraphics[width=\linewidth]{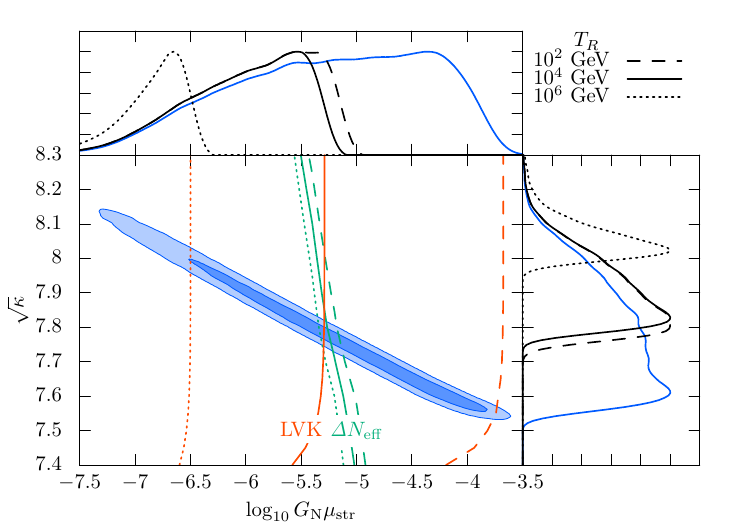}
        \caption{META-LS ($\tilde{c}=0.23$)}
        \label{fig:META-LS1}
    \end{subfigure}
    \begin{subfigure}{0.45\textwidth}
        \centering\includegraphics[width=\linewidth]{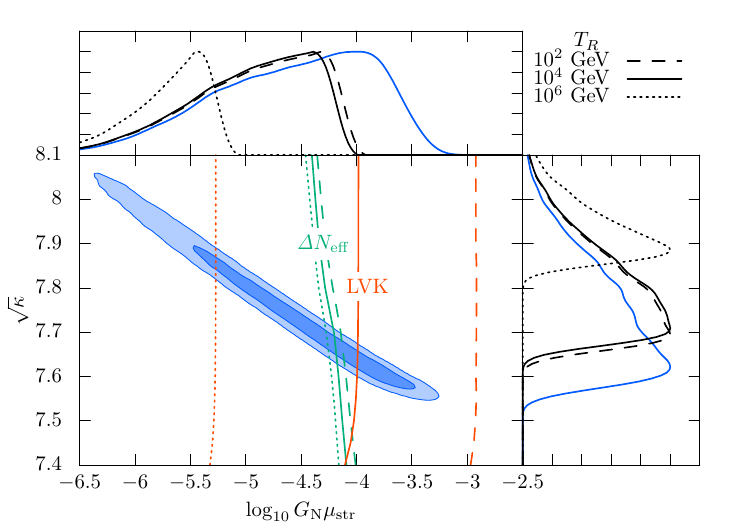}
        \caption{META-LS ($\tilde{c}=0.57$)}           
        \label{fig:META-LS2}
    \end{subfigure}
    \caption{Same as in \cref{fig:META-L}  but
     for META-LS which includes the GW contributions from the segments.}
    \label{fig:META-LS}
\end{figure}

In \cref{fig:META-LS}, we present the results for the META-LS scenario, where the energy loss from string segments primarily contributes to the GW spectrum. 
For the segment contribution, we sum up the mode up to $k_\mathrm{max}=10^5$ following Refs.\,\cite{Leblond:2009fq,Buchmuller:2021mbb}.
For $G_\mathrm{N}\mu_\mathrm{str} \simeq 10^{-5}$, 
the contribution from the segment 
is comparable to that from the cusps
in the GW spectrum in the PTA signal region. 
Accordingly, the required value of $G_\mathrm{N}\mu_\mathrm{str}$ is slightly lower than in the META-L scenario. 

There are several cautions regarding the GW contribution from the segments. 
First, since the strings in the present model carry $\uone_\mathrm{B-L}$ flux, 
the extent to which segments contribute to the GW energy density spectrum remains an open question. 
Additionally, the high-frequency behavior 
of the segment contributions is given by,
\begin{align}
    \Omega_\mathrm{GW}|_\mathrm{segment} 
    \propto \begin{cases}
        f^0&(f\ll k_\mathrm{max}\times f_*)\ ,\\ 
         f^{-1}& (f \gg k_\mathrm{max}\times f_*)\ ,
    \end{cases}
\end{align}
which can be read off from  Ref.\,\cite{Buchmuller:2021mbb}.
Here, $f_*$ denotes the characteristic 
frequency.%
\footnote{For the GW contributions from the segments from the long strings, for example, 
the fall off behavior in Ref.\,\cite{Buchmuller:2021mbb}
can be reproduced by taking $f_* = a(t_s)/a(t_0)\times\femit$ with $\femit =\order{10^{2}}\times(\Gamma G_\mathrm{N}\mu_\mathrm{str})^{-3/4}\times t_s^{-1}$ and $t_s=\Gamma_d^{-1/2}$ for $k_\mathrm{max}=10^5$. We confirmed the same behavior in our numerical analysis. }
Note that 
the presence of the early MD era does not affect the above flat spectrum for $f<k_\mathrm{max}\times f_*$,
since the GW emission from the segments occurs after the string breaking. 
Therefore, if $k_\mathrm{max}$ is extremely large, the contribution from segments potentially conflicts with the LVK constraint.
The maximum harmonic number $k_\mathrm{max}$,
however, depends on how the GWs are emitted from the endpoint monopoles, which require further investigation.
In our analysis, the above issues are left for future work, and the contribution from segments is calculated following Ref.\,\cite{NANOGrav:2023hvm}, summing up to harmonic modes with $k_\mathrm{max} = 10^5$.
In this sense, the constraints on the META-LS shown in the figure should be regarded as a reference (see also discussion in Ref.\,\cite{Servant:2023tua}).

\begin{figure}
    \centering
        \centering
        \includegraphics[width=0.5\linewidth]{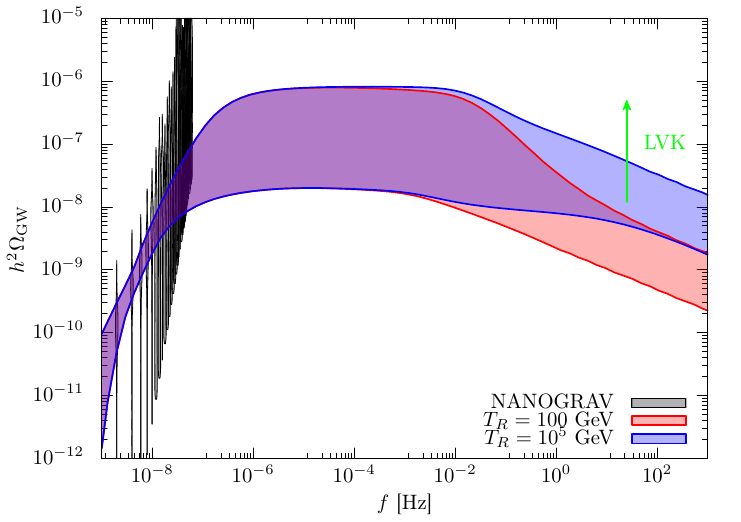} 
\caption{The GW energy density spectrum for the META-LS scenario is shown for $T_R = 10^5\,\mathrm{GeV}$ (blue) and $T_R = 10^2\,\mathrm{GeV}$ (red). The gray violin plots represent the reconstructed NANOGrav signal~\cite{NANOGrav:2023hvm}, while the green arrow indicates the LVK constraints on the power-law GW spectrum $f^{-1/3}$ at $25\,\mathrm{Hz}$~\cite{KAGRA:2021kbb}. The band width of each spectrum corresponds to the 95\% credible region of the posterior distribution for $(\log_{10}G_\mathrm{N}\mu_\mathrm{str}, \sqrt{\kappa})$, as shown in Fig.\,\ref{fig:META-LS1}. For the segment contribution, modes are summed up to $k_\mathrm{max} = 10^5$.
}
    \label{fig:spectrum}
\end{figure}

Finally, we present the GW spectrum for metastable string loops for $T_R = 10^5\,$GeV and $T_R = 10^2\,$GeV, as shown in Fig.\,\ref{fig:spectrum}.
The META-LS case is assumed with $k_\mathrm{max}=10^5$ for the segment contributions. The band width of each spectrum corresponds to the 90\% credible region of the posterior distribution of $(\log_{10}G_\mathrm{N}\mu_\mathrm{str}, \sqrt{\kappa})$ shown in Fig.\,\ref{fig:META-LS1}. 
The fall-off at low frequencies reflects the metastability of the strings, while the suppression at high frequencies arises from the early MD era associated with low reheating temperatures.

The figure demonstrates that the PTA signal can be well explained by GWs from metastable strings. In the META-LS scenario, the segment contribution is comparable to that of cusps in the GW spectrum within the PTA signal region for $G_\mathrm{N}\mu_\mathrm{str} \simeq 10^{-5}$. As a result, the GW spectrum in the META-L case is lower than in the META-LS case for the same $G_\mathrm{N}\mu_\mathrm{str}$. Additionally, the slope of the low-frequency spectrum is slightly flatter in the META-L case compared to the META-LS case~(see Ref.\,\cite{NANOGrav:2023hvm}).

\section{Conclusions and Discussions}
\label{sec:conclusions}
In this paper, we discussed a supersymmetric new inflation model that can generate a metastable string network through two-step symmetry breaking,  
$\SU(2)\to \uone_G \to \mbox{nothing}$.
In this model, the inflaton corresponds to the field 
responsible for the first step of two-step symmetry breaking.
Accordingly, the first symmetry breaking occurs at the onset of inflation, 
while the second breaking takes place after inflation.
This set up allows the formation of the string network without monopole production.
We also find that the model can successfully 
provide the GUT scale breaking scales 
favored by the metastable string interpretation
of the PTA GW signal.

We have also discussed the decay processes of the inflaton sector 
fields in detail.
Then, we found that the model requires rather low reheating temperature 
to avoid the gravitino overproduction from the decays 
of the inflaton sector fields.
Interestingly, this low reheating temperature enables the GW spectrum to be consistent with  LVK  constraints for a relatively high string tension  $G_{\mathrm{N}} \mu_\mathrm{str} \sim 10^{-5}$.
The characteristic stochastic GW spectrum predicted by this model
presented in \cref{fig:spectrum}
can be tested by future PTA experiments \cite{Carilli:2004nx, Janssen:2014dka, Weltman:2018zrl} and interferometer experiments \cite{Seto:2001qf, Kawamura:2006up, Yagi:2011wg, Isoyama:2018rjb, Crowder:2005nr, Corbin:2005ny, Harry:2006fi, Somiya:2011np, Aso:2013eba, KAGRA:2018plz, KAGRA:2019htd, Michimura:2019cvl, LIGOScientific:2016wof, Reitze:2019iox, Punturo:2010zz, Hild:2010id, Sathyaprakash:2012jk, ET:2019dnz} (see also Ref.\,\cite{Schmitz:2020syl} for theoretical study of the future sensitivities).

To provide a working example for the successful cosmology,
we also discussed the non-thermal leptogenesis as well as 
the LSP dark matter scenario.
Our model proved consistent with these scenarios with PeV-scale gravitino/sfermion masses and the anomaly-mediated gaugino masses in the TeV range.

Notice that several challenges remain unexplored.
Firstly, in this work, we assumed that the cosmic strings 
generated by the $\uone_G$ symmetry breaking are classically stable. 
However, since the $\uone_G$-breaking VEV is close to the $\SU(2)$-breaking VEV, it is non-trivial whether the static string configuration has no instabilities 
such as $W$-condensation~\cite{Ambjorn:1989sz}.
It is also important to explore the possibility of monopole-\modifiedat{JCAP2}{antimonopole} pair creation during string reconnection.

Besides, the cosmic strings considered here are much thicker than the inverse of the mass of the $\uone_G$ gauge boson (see \cref{fig:string_configuration}), requiring a more precise analysis of the string-breaking rate via monopole pair production. 
This would refine the approximation used in Ref.\,\cite{Preskill:1992ck}, which neglected the string thickness and monopole size (see, e.g., Refs.\,\cite{Shifman:2002yi,Chitose:2023dam}).
Furthermore, although we utilized the VOS model calibrated for Nambu-Goto strings, the string thickness may also influence the evolution of the cosmic string network.
These are left for future investigation.

The evolution of the string segments in the present model is 
also an open question, since the metastable string segments are associated with the $\uone_{B-L}$ magnetic field which is confined along the segments.
In particular, computation of the GW spectrum requires understanding of frictions from SM plasma through which monopoles lose energy.
Besides, such energy release through the confined $\uone_{B-L}$ magnetic flux 
at the cosmic temperature below $1$\,MeV could have detectable effects on the BBN and CMB.
We will study those effects of the $\uone_{B-L}$ magnetic field in future work.

Also, the post-inflation dynamics must be investigated more in detail.
As we discussed in \cref{sec:reheating}, 
the energy fractions of the inflaton sector fields in the early matter domination
are given by the ensemble average over many Hubble patches
since they are influenced by the Hubble fluctuations.
Since the gravitino abundance depends on them,
statistical analysis is required to narrow down the viable parameter space of this model.

\section*{Acknowledgments}
MI would like to thank F.~Takahashi for bringing our attention to the New Inflation model that induces symmetry breaking.
This work is supported by Grant-in-Aid for Scientific Research from the Ministry of Education, Culture, Sports, Science, and Technology (MEXT), Japan, 21H04471, 22K03615 (M.I.), 20H01895, 20H05860 and 21H00067 (S.S.) and by World Premier International Research Center Initiative (WPI), MEXT, Japan. 
The work of S.S. is supported by DAIKO FOUNDATION. 
This work is also supported by Grant-in-Aid for JSPS Research Fellow 
JP24KJ0832 (A.C.). 
This work is supported by FoPM, WINGS Program, the University of Tokyo (A.C.).
This work is supported by JST SPRING, Grant Number JPMJSP2108 (S.N.).

\appendix

\section{A Simple Example of Model with Large \texorpdfstring{$\boldsymbol{\lambda_X}$}{λX}}
\label{sec:largelam}
\begin{table}[t]
\centering
\begin{tabular}{c||c|c|c|c}
& SU(2) & $\mathbb{Z}_{12R}$ & $\mathbb{Z}_{6}$ &$\uone_H$ \\
\hline
$\phi^a$ & $\mathbf{3}$ & $2$ & $2$ &$0$\\
\hline
$H$ & $\mathbf{2}$ & $-1$ & $-1$ & $-1$ \\
\hline
$\bar{H}$ & $\bar{\mathbf{2}}$ & $1$ & $3$ &$+1$\\
\hline
$X$ & -- & $2$ &$0$ &$0$\\
\hline
$Y$ & -- & $2$ & $0$&$0$\\
\hline
$Z$ & -- & $-2$ &$2$&$0$ \\
\hline
$\bar{Z}$ & -- & $4$ & $-2$&$0$\\
\end{tabular}
    \caption{Charge assignment of the extended model for $n=3$ and $n'=1$ with $\mathbb{Z}_{12R}$ and $\mathbb{Z}_{6}$ symmetry.}   
    \label{tab:symmetry2}
\end{table}
To provide a large $\lambda_X$, let us consider 
the following model behind the superpotential 
in \cref{eq:Wpot},
\begin{align}
    W_{Z\bar{Z}}=\frac{1}{2} \lambda_Z Z(\phi\cdot \phi) + 
    \frac{1}{3!}\frac{1}{M_\mathrm{Pl}}\lambda_{\bar{Z}}X\bar{Z}^3 + M_{Z\bar{Z}} Z \bar{Z}+ Xv_X^2\ .
\end{align}
where $\lambda$'s are $\order{1}$ coefficients and $M_{Z\bar{Z}}$ is the mass of the pair of the chiral fields $Z$ and $\bar{Z}$.

By integrating out $Z$ and $\bar{Z}$, the above model leads to
\begin{align}
    W_\mathrm{eff} = X v_X^2 + \frac{\lambda_Z^3 \lambda_{\bar{Z}}}{8\cdot 3!} \frac{1}{M_{Z\bar{Z}}^3M_\mathrm{Pl}} (\phi\cdot\phi)^3\ ,
\end{align}
which realized the model with $n=3$.
In fact, the effective $\lambda_X$ is given by
\begin{align}
    \lambda_X = \frac{\lambda_Z^2\lambda_{\bar{Z}}}{8\cdot 3!} \frac{M_\mathrm{Pl}^3}{M_{Z\bar{Z}}^3}\ .
\end{align}

Note that the effective field theory given by 
K\"ahler potential given by the field expansion in powers of the inverse of $M_\mathrm{Pl}$ is not spoiled as long as
\begin{align}
    M_{Z\bar{Z}} \gtrsim \frac{M_\mathrm{Pl}}{4\pi}\ .
\end{align}
Thus, we find that the upper limit on the effective coupling is given by,
\begin{align}
    \lambda_X \lesssim \lambda_Z^2 \lambda_{\bar{Z}} 
    \times \frac{4\pi^3}{3} \ ,
\end{align}
which allows $\lambda_X = \order{10}$.
Therefore, the model with $n=3$ can be consistent with 
the SU(2) breaking scale consistent with NANOGrav.
Since $Z$'s have the mass of $\order{10^{17}}$\,GeV,
they do not affect the dynamics of inflation discussed in the previous section.

\section{Asymptotic Behavior of Gravitational Wave Spectrum}
\label{sec:GW}
In this appendix, we discuss the asymptotic behavior of the GW energy density spectrum for a large frequency $f$. The GW energy density spectrum is given by,
\begin{align}
\label{eq:OmegaGW}
    \Omega_\mathrm{GW}(f) &= 
    \frac{f}{\rho_c} 
    \dv{\rho_\mathrm{GW}(f,t_0)}{f}\ , \cr
    &= \sum_{k\ge 1}\frac{1}{\rho_c} \frac{2k}{f}    \frac{\sqrt{2}\mathcal{F}_\alpha P_k}{\gamma_\alpha\alpha(\alpha + \Gamma G_\mathrm{N} \mu_\mathrm{str} )}
    \int_{t_\mathrm{form}}^{t_0} dt \qty(\frac{a(t)}{a(t_0)})^{5}\qty(\frac{a(t_i)}{a(t)})^3
     \frac{C_\mathrm{eff}(t_i)}{t_i^4} \cr 
    &\phantom{XXXXXXXXXXXXXXXXXX}\times \Theta(t-t_i)\Theta(t_i - t_\mathrm{osc})
     \Theta(t_i - \ell_*/\alpha)
     \ .
\end{align}
The derivation of this expression follows Ref.\,\cite{Gouttenoire:2019kij}, and the notation for parameters is also generally consistent with the reference. The GW spectrum is obtained by integrating over the emission time $t$ from the formation time of the string network $t_\mathrm{form}$ to the present Universe $t_0$.%
\footnote{In the second step function, $t_\mathrm{osc}$ denotes the time when long strings begin to oscillate freely. The third step function ensures that the loop length is sufficiently large so that particle production from cusps to become negligible~\cite{Gouttenoire:2019kij}. However, these step functions are not relevant for discussing the asymptotic behavior addressed in this appendix.
}
The scale factor is represented by $a(t)$.

The parameter $\alpha$ represents the string loop length normalized by the cosmic time at the production $t_i$, such that $\ell = \alpha t_i$. The quantities $\mathcal{F}_\alpha$ and $\gamma_\alpha$ denote the fraction of loops and their corresponding boost factor, respectively, for loops formed with the length $\ell = \alpha t_i$. The loop formation efficiency, $C_\mathrm{eff}(t)$, is determined by solving the VOS equations for a given loop-production coefficient $\tilde{c}$, a parameter in the VOS model. In the scaling regime of the string network, $C_\mathrm{eff}(t)$ becomes a constant in time,
\begin{align}
\label{eq:Ceff}
    C_\mathrm{eff}|_{\tilde{c}=0.23} \simeq 
    \begin{cases}
    5.4   & p=4\,\mathrm{(RD)} \\
        0.39  &p=3\,\mathrm{(MD)} 
    \end{cases}\ ,
    \qquad
    C_\mathrm{eff}|_{\tilde{c}=0.57} \simeq 
    \begin{cases}
      1.3   & p=4\,\mathrm{(RD)} \\
        0.23  & p=3\,\mathrm{(MD)}
    \end{cases}\ ,
\end{align}
for the RD and MD eras, respectively. Here, the case $\tilde{c}\simeq 0.23$ corresponds to the VOS model calibrated to match Nambu-Goto simulations~\cite{Martins:2000cs} while $\tilde{c}\simeq 0.57$ to the VOS model calibrated to Abelian-Higgs simulations~\cite{Martins:2000cs}.

For a given GW frequency $f$ and harmonic mode $k \in \mathbb{N}_+$, the loop production time is given by,
\begin{align}
\label{eq:tprod}
    t_i(f,t) &= 
        \frac{1}{\alpha + \Gamma G_\mathrm{N} \mu_\mathrm{str} }\qty[\frac{2k}{f}\qty(\frac{a(\temit)}{a(t_0)}) + \Gamma G_\mathrm{N} \mu_\mathrm{str}\times \temit]\ . 
\end{align}
The frequency dependence in \cref{eq:OmegaGW} appears only through the combination $k/f$. The harmonic mode dependence, on the other hand, also appears through the GW power spectrum $P_k$,
\begin{align}
\label{eq:Pk}
    P_k = \frac{\Gamma G_\mathrm{N}\mu_\mathrm{str}^2} {\zeta_R(4/3)}k^{-4/3}\ , \quad \zeta_R(s)= \sum_{k\ge 1} k^{-s}\ ,
\end{align}
where we assume that the GW's from the string network are dominated by those from the cusps on the string loops~\,\cite{Blanco-Pillado:2017oxo}.

Now, let us consider the asymptotic behavior of the integrand of \cref{eq:OmegaGW}. In the RD era, it asymptotes to
\begin{align}
\lim_{t\to \infty} \qty(\frac{a(t)}{a(t_0)})^{5}\qty(\frac{a(t_i)}{a(t)})^3
   \frac{1}{t_i^4}\bigg|_\mathrm{RD} \propto  
   t^{-3/2}\ .
\end{align}
Besides, the $\temit$ integration is dominated by the contribution from the region satisfying
\begin{align}
\label{eq:temitRD}
    \frac{2k}{f}\qty(\frac{t_\mathrm{emit}}{t_0})^{1/2} \sim \Gamma G_\mathrm{N} \mu_\mathrm{str}\times t_\mathrm{emit}\ ,
\end{align}
and hence $t_\mathrm{emit} \propto (f/k)^{-2}$.
Accordingly, in the RD era, the
frequency dependence of the GW spectrum for a given harmonic mode $\Omega_{\mathrm{GW}}^{(k)}$ asymptotes to
\begin{align}
\label{eq:OmegaKRD}
    \Omega^{(k)}_{\mathrm{GW}}|_\mathrm{RD} \propto  \frac{k}{f} \times t_\mathrm{emit}^{-1/2}\times  P_k \propto \qty(\frac{k}{f})^0 \times P_k\ .
 \end{align} 
This explains why the GW spectrum forms a plateau in the high-frequency region as in \cref{eq:plateau}.

In the MD era, on the other hand, the integrand of \cref{eq:OmegaGW} asymptotes to
\begin{align}
\lim_{t\to \infty} \qty(\frac{a(t)}{a(t_0)})^{5}\qty(\frac{a(t_i)}{a(t)})^3
   \frac{1}{t_i^4}\bigg|_\mathrm{MD} \propto  
   t^{-2/3}\ .
\end{align}
Thus, the $\temit$ integration is dominated by the region where 
\begin{align}
      \frac{2k}{f}\qty(\frac{a(t_\mathrm{emit})}{a(t_0)}) \ll \Gamma G_\mathrm{N} \mu_\mathrm{str}\times t_\mathrm{emit} \ ,
\end{align}
and hence, the integrated value in \cref{eq:OmegaGW} becomes independent of $f$.
Thus, in the MD era, the frequency dependence of $\Omega_\mathrm{GW}^{(k)}$ asymptotes to
\begin{align}
\label{eq:OmegaKMD}
    \Omega^{(k)}_{\mathrm{GW}}|_\mathrm{MD} \propto  \frac{k}{f} \times  P_k \ .
 \end{align}

Finally, let us discuss the summation over the harmonic mode.
Consequently, the total differential GWs emitted in the early MD era is proportional to
\begin{align}
\label{eq:OmegaGWkmax}
\Omega_\mathrm{GW}|_\mathrm{MD} \propto    \frac{1}{f} \times \sum_{k=1}^{k_\mathrm{max}} k^{-1/3} \sim \frac{1}{f} \times k_\mathrm{max}^{2/3} \ ,
\end{align}
we have used \cref{eq:Pk}. 
By noting that the harmonic modes with $k$ satisfying
\begin{align}
    f > f^{(k)}(T_R)\ ,
\end{align}
are emitted in the MD era, 
the $k_\mathrm{max}$ in \cref{eq:OmegaGWkmax} 
is given by $k_*$, 
\begin{align}
   f =  f^{(k_*)}(T_R)\ .
\end{align}
Thus, since $k_*\propto f$, we find that the total GW spectrum is suppressed by
\begin{align}
\label{eq:OmegaGWMD}
    \Omega_\mathrm{GW}|_\mathrm{MD}  \propto f^{-1/3} \ , 
\end{align}
for $f \gg f^{(1)}(T_R)$.

\section{String Tension for \texorpdfstring{$\boldsymbol{\zeta=0}$}{ζ=0}}
\label{sec:tension}
In this Appendix, we present an approximate analytic formula for our cosmic strings in the small $\zeta$ limit.
Here, supergravity effects and non-minimal couplings in the Kähler potential are ignored.
Here, we specialize to $(n, n')=(3, 1)$. Also, we employ the unit system with $M_{\mathrm{Pl}}=1$.

Plugging the ansatzes \eqref{eq:string ansatz1}, \eqref{eq:string ansatzX}, \eqref{eq:string ansatzY} and \eqref{eq:string ansatz2} and $X\equiv Y\equiv 1$ into the action, we find the tension to be
\begin{equation}
    \mu_\text{str}=2\pi \int_0^\infty d\rho\,\rho \Big[ 
        \frac{2}{g^2\rho^2}(\partial_\rho a_s)^2
        +2\vone^2\pqty{(\partial_\rho h_s)^2+\frac{1}{\rho^2}a_s^2h_s^2} 
        +\lambda_Y^2 \vone^{12}(1-h_s^6)^2
    \Big]\ .\label{eq:simpletension}
\end{equation}
The equations of motion are
\begin{gather}
    \partial_\rho^2h_s+\frac{1}{\rho}\partial_\rho h_s-\frac{1}{\rho^2}a_s^2h_s+3M^2h_s^5(1-h_s^6)=0\ , \label{eq:heom}\\
    \partial_\rho^2 a_s-\frac{1}{\rho}\partial_\rho a_s -g^2\vone^2 a_sh_s^2=0\ ,\label{eq:aeom}
\end{gather}
where $M:=\lambda_Y\vone^5$.

Since the potential is suppressed by $M^2\propto M_{\mathrm{Pl}}^{-10}$, the situation is close to the type-I Abrikosov-Nielsen-Olesen string~\cite{Abrikosov:1956sx,Nielsen:1973cs}.
Thus, we may proceed by mimicking the argument by Yung~\cite{Yung:1999du}.
We suppose the gauge field is confined in a small radius $\rho_g$, whereas $h_s$ start to grow when $\rho>\rho_g$.
Our ansatz for $a_s$ is
\begin{equation}
    a_s(\rho)=\begin{cases}
        1-\frac{\rho^2}{\rho_g^2} & (\rho\ll \rho_g)\\
        0 & (\rho\gg \rho_g)
    \end{cases}
\end{equation}
The branch for small $\rho$ satisfies \cref{eq:aeom} with $h_s=0$.

Next, we solve \cref{eq:heom} for $\rho\gg \rho_g$, where $a_s=0$.
We employ different approximations in two regions of $\rho$ and glue them together.

For $h_s\ll 1$, the potential term can be neglected. The solution is
\begin{equation}
    h_s(\rho)=C\ln\frac{\rho}{\rho_g}\ ,
\end{equation}
where $C$ is an integral constant and we required that $h_s(\rho_g)=0$.
For this approximation to be valid,
\begin{equation}
    \partial_\rho^2 h_s\sim \frac{1}{\rho}\partial_\rho h_s\sim \frac{C}{\rho^2}\gg 3M^2h_s^5(1-h_s^6)>M^2, 
\end{equation}
or
\begin{equation}
    \rho^2\ll \frac{C}{M^2}
\end{equation}
is required.

For $h_s\sim 1$, $\tilde{h}_s:=1-h_s$ is small.
To its first order, \cref{eq:heom} reads
\begin{equation}
    \partial_\rho^2 \tilde{h}_s+\frac{1}{\rho}\tilde{h}_s-18M^2\tilde{h}_s=0\ .
\end{equation}
This can be solved as
\begin{equation}
    \tilde{h}_s(\rho)=BK_0(3\sqrt{2}M\rho)\ ,
\end{equation}
where $B$ is another integral constant and $K_0$ is the modified Bessel function of the second kind.
Note that $I_0$, the modified Bessel function of the first kind, is incompatible with the boundary condition at $\rho\to\infty$.
The $\order{\tilde{h}_s^2}$ term in \cref{eq:heom} is $-45M^2\tilde{h}_s^2$, and thus, the range of validity is $\tilde{h}_s<\order{0.1}$.

We must join the two at some $\rho=\rho_0$, where both the approximations are applicable.
We will see later that $\rho_0$ can be chosen so that $3\sqrt{2}M\rho_0 \lesssim \order{0.1}$.
Consequently, the asymptotic form for small $z$
\begin{equation}
    K_0(z)\simeq -\ln \frac{z}{2}-\gE
\end{equation}
is valid, where $\gE\approx 0.577$.
Requiring $h_s(\rho_0)$ and $h_s'(\rho_0)$ to agree on the both sides, we have
\begin{equation}
    B=C=\bqty{\ln\frac{\rho_M}{\rho_g}}^{-1}\ ,
\end{equation}
where
\begin{equation}
    \rho_M:=\frac{\sqrt{2}e^{-\gE}}{3M}\ .
\end{equation}

The contribution to $\mu_{\mathrm{str}}$ from $\rho<\rho_g$ is
\begin{equation}
    {\mu_{\mathrm{str}}}|_{\rho<\rho_g}=2\pi\int_0^{\rho_g}d\rho\,\rho \frac{2}{g^2\rho^2}(\partial_\rho a_s)^2=\frac{8\pi}{g^2\rho_g^2}\ . \label{eq:smalltension}
\end{equation}
The contribution from $\rho>\rho_g$ is
\begin{equation}
    {\mu_{\mathrm{str}}}|_{\rho>\rho_g}=2\pi\int_{\rho_g}^{\rho_0}d\rho\,\rho \bqty{2\vone(\partial_\rho h_s)^2+\vone^2M^2(1-h_s^6)}\simeq 4\pi \vone^2 \pqty{\ln\frac{\rho_M}{\rho_g}}^{-1}\ , \label{eq:largetension}
\end{equation}
which is saturated by the kinetic term of $h_s$.
Requiring $\rho_g$ to minimize the total tension, we get
\begin{equation}
    \frac{\rho_M}{\rho_g}\simeq \frac{g\vone\rho_M}{2}\pqty{\ln\frac{g\vone\rho_M}{2}}^{-1} \ .
\end{equation}
Substituting this into Eqs.~\eqref{eq:smalltension} and \eqref{eq:largetension}, we get the final result
\begin{equation}
    \mu_{\mathrm{str}}\simeq 4\pi \vone^2 \bqty{\ln( \frac{g\vone\rho_M}{2}\pqty{\ln\frac{g\vone\rho_M}{2}}^{-1})}^{-1}\ . \label{eq:approxtension}
\end{equation}

Substituting the values of BP$_1$ ($\vone=0.03$, $\lambda_Y=0.1$ and $g=1$), we have $\mu_{\mathrm{str}}/(2\pi \vone^2)\simeq0.170$, in reasonable agreement with the right side of \cref{fig:string_configuration}.
Also, choosing $3\sqrt{2}M\rho_0=0.35$ gives $\rho_0^2/(C/M^2)=0.080$ and $h_s(\rho_0)=0.90$, satisfying all the requirements.

\bibliographystyle{apsrev4-1}
\bibliography{bibtex}

\end{document}